*Title:*

Deciphering gene regulation from gene expression dynamics using deep neural network

*Author Affiliation:*

Jingxiang Shen[a,b,c], Mariela D. Petkova[d,1], Yuhai Tu[e], Feng Liu[a,b], and Chao Tang[a,b,c]

[a]Center for Quantitative Biology, [b]School of Physics, and [c]Peking-Tsinghua Center for Life Sciences, Peking University, Beijing 100871, China

[d]Joseph Henry Laboratories of Physics and Lewis-Sigler Institute for Integrative Genomics, Princeton NJ 08544

[e]IBM T. J. Watson Research Center, Yorktown Heights, New York 10598, USA.

[1]Present address: Program in Biophysics, Harvard University, Cambridge MA 02138

*Corresponding Author:* Chao Tang, Center for Quantitative Biology, Peking University, Beijing 100871, China. Phone: +86-10-62752003. Email: tangc@pku.edu.cn.

**Abstract**

Complex biological functions are carried out by the interaction of genes and proteins. Uncovering the gene regulation network behind a function is one of the central themes in biology. Typically, it involves extensive experiments of genetics, biochemistry and molecular biology. In this paper, we show that much of the inference task can be accomplished by a deep neural network (DNN), a form of machine learning or artificial intelligence. Specifically, the DNN learns from the dynamics of the gene expression. The learnt DNN behaves like an accurate simulator of the system, on which one can perform *in-silico* experiments to reveal the underlying gene network. We demonstrate the method with two examples: biochemical adaptation and the gap-gene patterning in fruit fly embryogenesis. In the first example, the DNN can successfully find the two basic network motifs for adaptation – the negative feedback and the incoherent feed-forward. In the second and much more complex example, the DNN can accurately predict behaviors of essentially all the mutants. Furthermore, the regulation network it uncovers is strikingly similar to the one inferred from experiments. In doing so, we develop methods for deciphering the gene regulation network hidden in the DNN "black box". Our interpretable DNN approach should have broad applications in genotype-phenotype mapping.

**Significance**

Complex biological functions are carried out by gene regulation networks. The mapping between gene network and function is a central theme in biology. The task usually involves extensive experiments with perturbations to the system (e.g. gene deletion). Here, we demonstrate that machine learning, or deep neural network (DNN), can help reveal the underlying gene regulation for a given function or phenotype with minimal perturbation data. Specifically, after training with wild-type gene expression dynamics data and a few mutant snapshots, the DNN learns to behave like an accurate simulator for the genetic system, which can be used to predict other mutants' behaviors. Furthermore, our DNN approach is biochemically interpretable, which helps uncover possible gene regulatory mechanisms underlying the observed phenotypic behaviors.

**Introduction**

Complex biological functions are carried out by gene regulation networks. Uncovering the gene regulation network behind a function is one of the central themes in biology. Traditionally, this task usually involves extensive genetic and biochemical experiments with perturbations to the biological system. For example, in the classical gene knockout experiments, by observing the expression of gene *a* increasing (decreasing) when deleting gene *b,* one may infer that gene *a* is repressed (activated) by gene *b*. More recently, statistical and bioinformatical methods have been used to help mapping out genetic and protein interactions. These methods can be very powerful especially in analyzing high throughput experimental data and extracting information about correlations among the genes and proteins (1, 2). Another computational approach is "reverse engineering", that is to search computationally for possible regulation networks capable of executing a given function or generating a given phenotype. This approach has been widely applied to a variety of problems, including biochemical adaptation (3), noise reduction (4), cell cycle control (5), cell polarization (6), and fruit fly embryogenesis (7-10). However, while reverse engineering has been very effective and fruitful in dealing with small networks and functional modules with relatively low complexity, it faces certain fundamental challenges in scaling up to larger and more complex systems. First, the search space increases drastically with the network size, which could easily reach the limit of the available computation power. Furthermore, there is often a trade-off on model complexity. On one hand, the representation power of simple models (e.g. with few parameters or prefixed regulation forms) may not capture the intricate nonlinearity and the complex regulations of the system. On the other hand, complicated models (e.g. detailed differential equations) would have too many parameters for large systems, making them difficult to tune without overfitting.

We reason that these challenges may be met by using a deep neural network (DNN) (11) to simulate the system under study. First, DNN is a flexible framework covering a very large model space with many possibilities. Hence, we do not need to be constrained by details (at least in the beginning) such as the validity of predefined regulation forms and types, the way to describe cooperativity and regulation logic, the form of nonlinearity and the number of parameters in the equations, etc. Second, fitting a DNN model to data can be achieved very efficiently by the standard back-propagation algorithm, making it computationally tractable compared with other fitting, searching and optimization schemes. Third, for reasons not yet completely clear, although DNN seems to have a huge fitting power, it hardly overfits even without any special regularization techniques (12, 13). In a sense, DNN is a kind of "*implicit self-adapting model*", adjusting its own capacity to fit but not overfit.

However, there is no free lunch. There are new challenges in using DNN to decipher gene network. It is unclear whether one could successfully train a DNN-based model for the inverse modelling purpose. Biological data are often noisy, incomplete and less quantitative. Data from different sources may vary and can even be "inconsistent" with each other. Furthermore, even if one manages to obtain an accurate DNN-based simulator of the target system, another challenge is to interpret this over-parameterized black box (DNN) to explicitly extract the underlying gene regulation network. So far, nearly all the biological applications of DNN, such as those in bioinformatics (14, 15), focus on prediction rather than interpretation for the underlying mechanism. Although there are some studies on interpretable DNN with results in understanding the features learnt by feed-forward convolutional neural networks (16, 17), these methods cannot be applied directly here to uncover the gene network.

In this paper, we implement a recurrent architecture of DNN to simulate the dynamics (time evolution) of the gene network system and use the data for gene expression dynamics for training, forcing the DNN to learn the mechanism (causation) of the regulations. We further develop methods to manipulate the DNN black box to uncover the underlying biochemically meaningful regulation network. We first illustrate our method with a relatively simple example, biochemical adaptation, and show that it can successfully identify the well-known key regulation networks capable of adaptation. We then apply our method to a much more complex example with real data, the gap-gene patterning in fruit fly embryogenesis, where gene expression varies not only in time but also in space. We show that the DNN, trained with the wild type (WT) data in gene expression dynamics plus a couple of mutant snapshots, can accurately predict the behavior of essentially all the other mutants. Furthermore, we develop a simple method to interpret the DNN-based model, which reveals an underlying gene regulation network that is strikingly similar to the one inferred from other independent experiments. In the discussion section and the corresponding SI, we give some remarks on how one should make use of and expect from such a DNN-based model.

## Results

**Biochemical Adaption.**

We first demonstrate the basic idea and the implementation of our method with biochemical adaptation, a ubiquitous and well-studied cellular function. Regulatory network topologies for this function have been exhaustively studied (3). For simplicity, we consider the case in which the system consists of two genes (or two coarse-grained nodes). The time evolution of the system can be described by the following equation

$$g_i(t + dt) = g_i(t) + f_i(g_1, g_2, I)dt - \gamma_i g_i(t)dt,$$

where $g_i$ is the expression level of gene $i$ ($i$=1,2), $I$ the input signal, $f_i$ the synthesis rate of gene $i$, and $\gamma_i$ the degradation rate. For convenience, we write the above equation in a more compact form

$$\boldsymbol{g}(t + dt) = \boldsymbol{g}(t) + \boldsymbol{f}(\boldsymbol{g}, I)dt - \gamma \boldsymbol{g}(t)dt$$

where $\boldsymbol{g}$ and $\boldsymbol{f}$ are two component vectors and the same degradation rate $\gamma$ is used for both genes

(In principle, any degradation terms can be absorbed into the *f* function. Here we explicitly add a uniform "degradation rate" mainly to keep ***g*** from diverging, and also for setting an inherent time scale.) The *f* function contains information about gene-gene interactions, which is the target of learning by DNN. As illustrated in Fig. 1A, we use a DNN (see Methods) to generate the *f* function. The DNN uses the current value of ***g***(t) and *I*(t) as its input and generates *f* as its output. Time evolution of the genetic system then corresponds to the recurrent iteration of the DNN.

The learning of the *f* function is through training the DNN. A Loss function is defined as the Euclidian distance between the target and model trajectories

$$Loss = \sqrt{\sum_t w_t \big(\boldsymbol{g}(t) - \hat{\boldsymbol{g}}(t)\big)^2},$$

where $\hat{\boldsymbol{g}}(t)$ is the desired trajectory or target and the weight *w* can be adjusted to implement specific training constrains (*SI Appendix*, S1). Note that at each time step, $\boldsymbol{g}(t)$ instead of $\hat{\boldsymbol{g}}(t)$ is used for calculating $\boldsymbol{g}(t + dt)$ when simulating the model trajectory.

For adaptation, we require one of the genes (***g***$_1$) to carry out the adaptation function, that is, it should respond to a change in input signal (stimulus) but returns to its pre-stimulus level even when the signal persists (Fig. 1B). The other gene has no preset functional constraints and can play a regulatory role. The target can be chosen, for example, as a set of desired values for ***g***$_1$ at certain time points (Fig. 1B). The standard training process (minimizing the Loss function by adjusting the weights in DNN) is described in Methods.

Training converges quickly (see Methods), yielding perfect adaptation (Fig. 1B). For this low-dimensional dynamical system, the trained *f* function can be plotted directly (see Fig. 1C for typical cross sections and *SI Appendix*, Fig. S1 for more details). Note that the *f* function is rather smooth and monotonic, which is indicative that the DNN is not overfitting. By observing whether $f_1$ and $f_2$ are increasing or decreasing with ***g***$_1$, ***g***$_2$ and input *I*, one can easily find the regulatory logic hiding in the thousands of DNN parameters. For example, $f_1$ increases with *I* (Fig. 1C, left panel), implying that $I$ activates $g_1$. From this information, one can construct the underlying gene network (Fig. 1D). The network consists of both an incoherent feed-forward loop and a feed-back loop, both known as the elementary motifs for adaptation (3, 18).

**Uncovering regulatory logic by *in-silico* mutation.**

For systems with more genes, which correspond to dynamical systems of higher dimensions, visualizing the $f$ function can be difficult and informative low-dimensional cross sections (such as Fig. 1C) may be hard to get. In order to uncover the regulatory logics of more complex gene networks and make the DNN-based model interpretable, we introduce a simple technique reminiscent of mutagenesis experiments in biology: blocking the regulation of gene 1 on gene 2, which experimentally would correspond to mutating the binding sites of transcription factor 1 on the regulatory region of gene 2. After the DNN has been trained, we can carry out the mutagenesis *in-silico* and observe the effect on gene expression: block the influence of $g_2$ on $f_1$, simulate the dynamics and see whether $g_1$ level increases or decreases compared with the "wild type". An increase in $g_1$ level in this "mutant" would imply a negative regulation or inhibition of $g_1$ by $g_2$; while a decrease means a positive regulation or activation. This "link blocking" mutation can be implemented by setting the value of $g_2$ to zero when and only when calculating $f_1$. In other words, the same trained DNN would run twice at each time step, calculating $f_1$ and $f_2$ separately (which is the implementation used here). Or, one can use a modified architecture, in which each output node ($f_1$ and $f_2$) has its own "private" DNN. To simulate the above mutation, train the system with full connection and then remove the link from $g_2$ to the DNN for $f_1$ (Fig. 2A). These two implementations generally do not make any difference for the *in-silico* mutation experiments. However, the second implementation can also be used for another purpose as discussed in the next paragraph. Fig. 2B shows two examples of such mutations. Performing such link blocking on each and all possible interactions yields a regulation network identical to Fig. 1D which is obtained by direct plot of the $f$ function.

The network obtained from the DNN-based model as shown in Fig. 1D contains two elementary adaptation motifs: negative feedback and incoherent feed-forward, each one of which would suffice for adaptation (3). This kind of "redundancy" is typical for the regulation network learnt by DNN without any constraints on the network topology. DNN simply searches for a solution that works rather than being minimal or understandable. Simple regularization (weight decay) (19, 20) helps little in sparsening the resulting regulation network, as parameters in DNN do not have explicit correspondence to edges in the regulation network (though it can make the $f$ function smoother) (see *SI Appendix*, S2 for results). Efficient reduction of the redundant links can be achieved by applying the "link blocking" technique introduced in the previous paragraph (Fig. 2A). By removing certain links in the beginning, the DNN is constrained to find a solution without those links in the regulation network. The idea is similar to dropout (21) but at the level of *regulation topologies* instead of DNN weights. In this way, we can find the minimal adaptive network topologies corresponding to the feed-back and the incoherent feed forward loops (Fig. 2C and *SI Appendix* Fig. S1).

With no prior knowledge on network connectivity, the DNN can also help us find minimal core networks for the function via a "greedy" iterative algorithm. Specifically, starting with a redundantly connected solution, block every single existing link and re-train the model, select the "offspring" that achieves adaptation the best, and iterate. A trajectory of regulation networks with increasing number of deleted links can thus be obtained (Fig. 2D). The minimum incoherent feed forward network emerges before the network has too few links to achieve adaptation.

**Gap-gene Patterning.**

Real regulation networks often have far more complex structures than the preceding simple example. Dynamics are often partially observable; noise and distortions in the measurement always exist. These features make deciphering regulations from real-world data much more challenging. However, real-world examples can provide plenty of specific situations which would help us learn to meet the challenges and to have a better understanding on what DNN based models can or cannot do. The example we study here is the gap-gene patterning in *Drosophila* embryogenesis, the first step to generate anterior-posterior (A-P) gene expression patterns from maternal morphogen gradients. This is one of the most thoroughly studied gene regulation networks for embryonic pattern formation and large amounts of data and knowledge have accumulated in the literature (22-26). These data and knowledge can be very helpful for model validation and interpretation.

In the gap gene patterning process, several maternal morphogens (Bicoid (Bcd), Cad, and Tor) form long-range gradients, influencing the later gap-gene dynamics by both providing them with spatially dependent external input and setting their initial conditions (the maternal component of Hunchback (mHb) is shaped by repression from the maternal effect protein Nanos (Nos) (27, 28)). Patterns of these morphogen gradients are taken to be fixed in our model as system's input (see *SI Appendix*, Fig. S3 for details). Gap-genes (*hunchback* (*hb*), *Krüppel* (*Kr*), *knirps* (*kni*), and *giant* (*gt*)) respond to the morphogen input and cross regulate each other, generating expression patterns as overlapping broad domains along the A-P axis at the blastoderm stage (Fig. 3A). The diffusion of gap proteins is ignored here (adding diffusion does not change the outcome; see *SI Appendix*, S3). So, gap-gene dynamics at each spatial point $x$ is calculated independently (Fig. 3B). This means how space is discretized does not matter much (we take 64 grids for simplicity). For each point $x$, the model has the same structure as Fig. 1A (or Fig. 2A), except that $\boldsymbol{g}$ and $\boldsymbol{f}$ both have four dimensions (Hb, Kr, Kni, Gt), and the input $\boldsymbol{I}$ has three (Bcd, Cad, Tor). (Note that for convenience at most times we do not distinguish between genes and gene products, so the notations like Hb and *hb* are used interchangeably). Time resolution have little effect on results, and is set around 100 seconds, about 1/10 of the typical timescale of the protein pattern change.

To train the model, we use a time series of wildtype (WT) gap-gene patterns (first 40 minutes of nucleus cycle (n.c.) 14, protein patterns), a snapshot pattern (at ~40 min. n.c. 14) of a mutant lacking all maternal inputs ($bcd^-nos^-tor^-$), and a mutant lacking all maternal inputs and maternal Hb ($bcd^-nos^-tor^-mhb^-$, snapshot pattern). These two mutant patterns help greatly in shaping the solutions; without them, roles of different maternal inputs tend to entangle with each other in the model, losing one-to-one correspondence with the real morphogens (see *SI Appendix*, Fig. S8B for details). It should be noted that certain detailed features of gap gene expression are known to be caused by regulations outside the gap-gene network, and are largely decoupled from the gap-gene system. These features are removed manually from the experimentally measured profiles (Fig. 3A and *SI Appendix*, S3).

The model is trained with conventional back-propagation algorithm and training converges very quickly (see Methods and *SI Appendix*, S4). The fitting error to target patterns (time series of normalized expression levels) is less than 1% (Fig. 3C; *SI Appendix*, Fig. S4A).

The trained DNN behaves almost exactly like a WT embryo in gap gene patterning. To see if the model really captures features of true regulation mechanisms, we perform *in-silico* mutations and compare the resulting patterns with the real mutants in the literature. It is straightforward to knockout genes in the model by just setting the variable(s) corresponding to the gene(s) to zero. Surprisingly, the model can "predict" the behavior for the vast majority of the mutants in the sense that the patterns of the model mutant and the corresponding real mutant (29, 30) are very similar (see Fig. 4A for examples and *SI Appendix*, Fig. S4B for an extensive list). To measure such similarity, note that the important and robust features of gap gene patterning are the positions of the gene expression peaks and boundaries along the A-P axis (31, 32). A quantitative evaluation can be carried out in this spirit. Features (peaks and boundaries) in these patterns are extracted and matched between the predicted and data profiles (Fig. 4B; see *SI Appendix*, S4 for details). 152 out of 166 features in 13 different real mutant patterns are successfully predicted by the model (8.4% missed), and 152 out of 159 predicted features are correct (4.4% false positive).

For those features correctly predicted, the root-mean-square error (RMSE) between the predicted and data positions is 4% embryo length (EL). This high level of accuracy is encouraging. To our knowledge no other differential-equation-based models in the literature achieved such a prediction power.

To place our model on a more stringent test, we put it against some of the "weird" mutants with double or triple mutations without tuning any trained parameters (Fig. 4C-F). There are only semi-quantitative data available for them, but all have interesting biological arguments behind (32, 33). For example, the posterior Gt domain is long thought to be strongly inhibited by Kr, but in $Kr^-$ mutant, it does not expand much towards the anterior pole (31). It could be possible that such expansion is stopped by the central Hb boundary, as posterior Gt domain is also known to be repressed by Hb (31). However, in $bcd^-Kr^-$ mutant where both the Kr domain and the central Hb boundary are lacking, Gt domain still does not expand much towards anterior (Fig. 4C, lower panel). It turned out that the anterior boundary of the posterior Gt domain is set mainly by the maternal Hb in the very early stage, and later refined by the inhibitions mentioned above (32). Remarkably, this nontrivial phenomenon is predicted by our model (Fig. 4C, upper panel). Similarly, the expression patterns of the mutants shown in Fig. 4D-F can also be predicted.

**Revealing the underlying regulation network.**
Those successful predictions on mutant patterns suggest that many essential characteristics of the gene regulation in this system are captured by our DNN. We proceed to uncover the underlying regulation network. As the model dimension is higher than the adaptation model, we use the "link blocking" method (Fig. 2A-B) to map out the regulations, which worked well for the adaptation example. (The method is further tested in an artificial example of spatial patterning, see *SI Appendix*, Fig. S10E). For example, we can block the influence from Gt to *kni* in our trained model (equivalent to experimentally mutating the binding sites of Gt on the regulatory region of *kni*). If as a result, Kni domain is found to expand toward Gt domain, we can conclude that Gt inhibits *kni* directly (Fig. 5A). "Directly" means that it is not mediated by other nodes within the model, as all edges apart from Gt-to-*kni* remains unchanged. Note that the anterior and posterior expression domains of Hb and Gt are known to be regulated differently, by different regulation modules (23, 31). So, in order to have a more direct comparison with biological knowledge, they are treated separately in the following text (denoted as $Hb_a$, $Hb_p$, $Gt_a$, and $Gt_p$).

The regulation network obtained by this method is shown in Fig. 5B (see also *SI Appendix*, Fig. S5 for comparison with other methods). Comparing it to the "real" gap-gene network inferred from experiments over the past decades (Fig. 5C), there are inspiring similarities and some interesting differences. First, inhibitory cross regulations among the gap-genes turned out to be identical for the model and "real" networks. In particular, the mutual inhibition between alternating domains ($Hb_a$, Kni, $Hb_p$; $Gt_a$, Kr, $Gt_p$) and the one-side inhibition between adjacent domains from posterior to anterior ($Hb_p$ to $Gt_p$ to Kni to Kr to $Hb_a$) are both present in our model and the "real" networks. This regulation structure makes much sense biologically, as it not only is the core motif for *Drosophila* gap-gene network (23), but also seems to play a pivotal role in long-germband insect evolution (23, 34-36). This core topology is captured by DNN, perhaps because it is implicated as the anterior shift of the abdomen domains in wild type gap-gene dynamics (37). Such shift is not an artifact or coincidence, but wide spread among different species (23, 34, 35).

Second, some "fake" (compared with the "real" network) activations emerged between gap genes. For example, Kr activates Kni in Fig. 5B but not in Fig. 5C. This type of "confusion" is quite understandable, as Kr inhibits Gt and Gt inhibits Kni, which could introduce an effective activation from Kr to Kni. Human researchers in the early days may also have confusions between double inhibition and direct activation (38). These fake activations could be regarded as sloppy modes in network topology, in that network function stays equivalent while structure change in certain ways. More pronounced equivalence of this type appears in the regulations of Bcd and Cad on anterior domains of Hb and Gt. Biologically, these domains are activated directly by Bcd and not affected by Cad (39, 40) as shown in Fig. 5C. But in Fig. 5B, the activation from Bcd is replaced by inhibition from Cad, which itself is inhibited by Bcd. Direct activation is replaced by double inhibition.

To see if these fake activations (between adjacent domains, from anterior to posterior) can be removed, we can train the model with these undesired interactions blocked by following the strategy developed in the adaptation example (Fig. 2C). Specifically, regulations from Cad to anterior Hb and Gt are also blocked at the same time, which forces Bcd to take on a direct activation role. Training under these constraints results in a gap-gene network nearly identical to the biological one (Fig. 6A) (apart from some self-activations, see *SI Appendix*, S5 for a discussion). This solution also gives fairly good predictions on mutant patterns. Using the criteria of Fig. 4B, 149 out of 166 features are predicted successfully, with positional RMSE of 3.9% EL, and 9 out of 158 predicted features are false positives.

Using similar ideas as those implemented before in adaptation (Fig. 2D), network sparsening can be carried out by link blocking. We first train the fully-connected gap-gene network, find the regulation link that has the least influence to pattern formation within the current solution (defined as having the minimum phenotype upon blockage), remove it, retrain the model and iterate. It is similar to the idea of leaning both architecture and weights out of training (41). Gap-gene cross regulations are thus being deleted one by one, causing prediction accuracy to drop along this trajectory (Fig. 6B). It should be noted that connectivity from the maternal inputs (thus positional information) are kept unchanged, so even all gap-gene interactions are removed, prediction accuracy does not drop too much. (See *SI Appendix*, Fig. S7 for results when all links, including those from maternal inputs, are dropped one by one). This result is reminiscent of the long-held claim that interpreting maternal morphogen sets the backbone structure of the gap-gene patterns, while cross regulations refine them (9, 23). On the way of this trajectory from over-connected to over-simplified solutions, there exists a point where redundant activations are removed while core regulation topology is basically kept (Fig. 6B).

**Discussion**

In this paper we explored the possibility of reverse engineering gene regulation networks with DNN. The DNN is first trained to simulate the system and to make accurate perditions. A "link blocking" method is developed to interpret the DNN-based model ("black box") and uncover the underlying regulation network. Our approach is demonstrated with two examples of very different complexity. It should have broad applications in genotype-phenotype mapping and in designing synthetic gene networks for desired functions.

In a purely technical sense, our approach could be viewed as a multi-layered version of the gene-circuit model (9), but the motivation and thus results are inherently different. Instead of seeking for a regulation network encoded explicitly as model parameters, we first train the model to be an accurate simulator of the target dynamics, using a flexible but over-parameterized black box (DNN). We then manage to obtain some interpretations on the trained model.

Over the past year, various ideas of simulating dynamical systems with DNNs have been proposed. Attentions of those researchers are mainly focused on the concepts itself, constructing model architectures informed of basic physics laws or dealing with sophisticated spatial differential operators (42-44), etc. Algorithms proposed in these literatures were typically demonstrated on synthesized data rather than real-world situations. Our attempts aim at exploring predictivity as well as interpretability, highlighting the abilities and limits for this type of approach in tackling tasks of real-world complexity. Both examples we used here are among the most thoroughly studied cases in biology. This knowledge turned out to be very helpful in understanding what the DNN is actually doing, as well as in what situations it may become unreliable. Based on the results and lessons we learned during this work, we now try to give some general remarks on how one should make use of, and expect from our method.

**Choice and effects of training data.**

Lack of data preprocessing (background removal, spatial-temporal alignment, etc.) would usually result in inconsistency within training data. Such inconsistency would largely cancel each other for a common deep learning task with large amounts of training data (45). However, in our case where independent training data is limited, including more mutant patterns for training usually brings more noise than informative constraints (*SI Appendix*, Fig. S8A). This is the reason why we only use WT and two pattern-less mutants profiles for training. On the other hand, insufficient training data would result in the entanglement of regulatory functions among different nodes, just like that of direct Bcd activation versus the double inhibition through Cad discussed above. Interestingly, however, it does not matter at all in situations where these nodes only need to function as a whole. For example, if only WT data is used for training, even the roles of Nos would entangle with Bcd and Cad, leading to bad predictions on maternal mutants where these factors operate alone. But when predicting gap-gene mutants, all maternal morphogens function as a whole and the predictions are not affected (*SI Appendix*, Fig. S8B).

**Sloppy modes in network topology.**

Network topology not constrained by the training data can result in the functions of certain genes to be entangled in the model (losing one-to-one correspondence from nodes in model to genes in reality), but they are still fully functional as a whole. We suggest that these unconstrained modes are ubiquitous in modelling complex systems and whether or not they will affect prediction depends on the specific prediction itself. Borrowing the concept of *parameter sloppiness* discussed in Ref. (46, 47), these modes can be viewed as sloppy modes in network topology. Similar to Ref. (46, 47), these sloppy modes involve multiple links in some redundant way. In the case of gap-gene system, none of the links lies purely along "rigid" directions and is absolutely irreplaceable. The network can always find its way to fit data and make correct predictions without each one of its links (for an enumeration, see *SI Appendix*, Fig. S6). So, in the gap-gene example (Fig. 5B), though the core motif can be successfully captured, these sloppy modes make it doubtful to trust every and all the links learnt by the DNN. Although there are ways to sparsen the regulation network (reducing the sloppy modes) (Fig. 6B), a more reliable way to reduce this redundancy is to constrain the network connection using prior (biological) knowledge (Fig. 6A).

**Insensitivity to missing genes.**

The DNN based model can be insensitive to missing genes, whose roles can be effectively absorbed into the rest of the network. For example, a 3-gap-gene model, lacking *kni*, can also fit and predict patterns of the three remaining genes quite well. The resulting regulation network is also similar to the "real" one (*SI Appendix*, Fig. S9). In fact, this may largely be the reason why the DNN works here: many unknown complexities are effectively absorbed. But such robustness can at the same time be unfavorable, for it may allow models with incorrect assumptions or lacking important components to slip through.

**Only robust mechanisms can be learned.**

DNN can be a powerful simulator for regulation networks. But it cannot be as successful for an *arbitrarily* synthesized dataset. In machine leaning, it has long been known that dataset should have some intrinsic structures to be "learnable". Here, whether a regulation network can be successfully simulated and reverse-engineered should depend on the network itself. We speculate that DNN can only learn "robust" mechanisms easily. To demonstrate in what case the DNN will fail, we take an oversimplified gap-gene network topology (*SI Appendix*, Fig. S10A, similar to that proposed in Ref. (48)) as an example. Taking this topology as the "ground truth", we implement it with Hill-functions, tune the parameters, and generate a set of synthesized WT and mutant patterns. Then we train a DNN model with exactly the same setting as Figs. 3 and 4. Interestingly, even with this "clean" synthesized data, the DNN fails to predict mutant patterns like $hb^-$ and $Kr^-$ (*SI Appendix*, Fig. S10B). The reason is straightforward: though able to generate a WT pattern similar to the real fruit fly, this artificial network, with too few links, breaks up if Hb or Kr is removed. Interestingly, the predictions from the learnt DNN for such deletions are closer to the real case than that of the artificial model – other expression domains still exist though shifted slightly, which is reminiscent of what happens for the real fruit fly. In a sense, it seems that the DNN possesses inherently certain "structure" that resembles the biology more than the oversimplified and overfitted "artificial ground truth", thus being able to guess out the former but not the latter from limited data. This structure may relate to what it usually refers as robustness. So, it may be risky to validate network-inferring algorithms only with the datasets synthesized arbitrarily.

**Methods**

**DNN Architecture.**

DNN for the adaptation task is a 3-layered fully connected network, consists an input layer with three nodes (representing $g_1$, $g_2$, and input *I*), two hidden layers each with 32 nodes, and an output node with two nodes ($f_1$ and $f_2$). The same network architecture is used for the gap gene system in Fig. 3-5, though the number of nodes in the input and output layers are changed to seven (4 gap proteins, and 3 maternal morphogens) and four (4 gap genes) to fit in the problem. A slightly different network architecture is used for Fig. 6, where there are six output nodes ($Hb_a$, $Hb_p$, Kr, Kni, $Gt_a$, $Gt_p$) in total, and each output node is calculated with a two-layered fully-connected DNN, with 7 input and 32 hidden nodes. ReLU is used for activation function except for the output layer, where the sigmoid function is used to keep its value (synthesis rate) bounded between 0 and 1.

Depth, width, and architecture of the DNN do not have much influence on this task. Different architectures all give basically the same results (*SI Appendix*, S7). Model capacity is no longer a limiting factor; and obviously, overfitting is not related to number of parameters for DNNs.

These networks are used to calculate gene expression state of *t+dt* from state *t*, and are iterated to calculate the whole dynamics. The reason for using this "primitive" form of recurrent network rather than more modern and standard ones like the LSTM, is that LSTM is designed to have long-term memory. But here we would like to assume the model to be Markovian. All "memories" passing from the current timepoint to the next are the gene expression levels, which are directly constrained by supervised training. The reasons are: (1) our task is easy to learn even without complicated long-term memories; (2) long-term dependence will bring much difficulty for interpreting and constraining the model, as regulations are no longer something between genes, but a fuzzy interconnected network connecting to all historical expression levels. In a sense, this temporal Markovian structure is also a vital constraint to tame the otherwise enigmatic DNN.

**Training.**

The DNN is trained by standard gradient-based optimization (Adam optimizer in Tensorflow, learning rate 0.001). No minibatch is used, as the training set is so small (only 9 frames for the gap gene network). Noise are introduced explicitly by two means: first, a Gaussian white noise is added directly to the pattern each time when iterating the differential equations, simulating the noise that have to be overcome during development. Second, an explicit Langevin noise is added to all the DNN parameters each time they update. A very weak weight decay is also added. See *SI Appendix*, S1 and S4 for detailed settings of the Loss function for each of the two tasks. Training converges usually before 8,000 steps, around two minutes on a desktop computer.

**Acknowledgements**

We thank Xiaojing Yang, Ning Yang, and Xiao Li for helpful discussions. The work was supported by the Chinese Ministry of Science and Technology (Grant No.2015CB910300) and the National Natural Science Foundation of China (Grant No. 91430217).

# References

1. R. De Smet, K. Marchal, Advantages and limitations of current network inference methods. *Nat. Rev. Microbiol.* **8**, 717-729 (2010).
2. M. Hecker, S. Lambeck, S. Toepfer, E. van Someren, R. Guthke, Gene regulatory network inference: data integration in dynamic models-a review. *Biosystems* **96**, 86-103 (2009).
3. W. Ma, A. Trusina, H. El-Samad, W. A. Lim, C. Tang, Defining network topologies that can achieve biochemical adaptation. *Cell* **138**, 760-773 (2009).
4. G. Hornung, N. Barkai, Noise propagation and signaling sensitivity in biological networks: a role for positive feedback. *PLoS Comput. Biol.* **4**, e8 (2008).
5. G. Yao, C. Tan, M. West, J. R. Nevins, L. You, Origin of bistability underlying mammalian cell cycle entry. *Mol. Syst. Biol.* **7**, 485 (2011).
6. A. H. Chau, J. M. Walter, J. Gerardin, C. Tang, W. A. Lim, Designing synthetic regulatory networks capable of self-organizing cell polarization. *Cell* **151**, 320-332 (2012).
7. L. Sánchez, D. Thieffry, A logical analysis of the *Drosophila* gap-gene system. *J. Theor. Biol.* **211**, 115-141 (2001).
8. E. Mjolsness, D. H. Sharp, J. Reinitz, A connectionist model of development. *J. Theor. Biol.* **152**, 429-453 (1991).
9. J. Jaeger *et al.*, Dynamic control of positional information in the early *Drosophila* embryo. *Nature* **430**, 368-371 (2004).
10. W. Ma, L. Lai, Q. Ouyang, C. Tang, Robustness and modular design of the *Drosophila* segment polarity network. *Mol. Syst. Biol.* **2**, 70 (2006).
11. Y. LeCun, Y. Bengio, G. Hinton, Deep learning. *Nature* **521**, 436-444 (2015).
12. J. Schmidhuber, Deep learning in neural networks: an overview. *Neural Netw.* **61**, 85-117 (2015).
13. C. Zhang, S. Bengio, M. Hardt, B. Recht, O. Vinyals, Understanding deep learning requires rethinking generalization. arXiv:1611.03530 (01 November 2016).
14. C. Angermueller, T. Parnamaa, L. Parts, O. Stegle, Deep learning for computational biology. *Mol. Syst. Biol.* **12** (2016).
15. S. Min, B. Lee, S. Yoon, Deep learning in bioinformatics. *Brief. Bioinform.* **18**, 851-869 (2017).
16. T. D. Kulkarni, W. F. Whitney, P. Kohli, J. B. Tenenbaum, Deep convolutional inverse graphics network. arXiv:1503.03167 (11 March 2015).
17. R. R. Selvaraju *et al.*, Grad-CAM: visual explanations from deep networks via gradient-based localization. arXiv:1610.02391 (07 October 2016).
18. W. Shi, W. Ma, L. Xiong, M. Zhang, C. Tang, Adaptation with transcriptional regulation. *Sci. Rep.* **7**, 42648 (2017).
19. S. J. Hanson, L. Y. Pratt (1988) Comparing biases for minimal network construction with back-propagation. in *NIPS'88*, pp 177-185.
20. A. Krogh, J. A. Hertz (1992) A simple weight decay can improve generalization. in *NIPS'92*, pp 950-957.
21. N. Srivastava, G. Hinton, A. Krizhevsky, I. Sutskever, R. Salakhutdinov, Dropout: a simple way to

prevent neural networks from overfitting. *J. Mach. Learn. Res.* **15**, 1929-1958 (2014).
22. J. O. Dubuis, R. Samanta, T. Gregor, Accurate measurements of dynamics and reproducibility in small genetic networks. *Mol. Syst. Biol.* **9**, 639 (2013).
23. J. Jaeger, The gap gene network. *Cell. Mol. Life. Sci.* **68**, 243-274 (2011).
24. A. Pisarev, E. Poustelnikova, M. Samsonova, J. Reinitz, FlyEx, the quantitative atlas on segmentation gene expression at cellular resolution. *Nucleic Acids Res.* **37**, D560-D566 (2009).
25. F. Liu, A. H. Morrison, T. Gregor, Dynamic interpretation of maternal inputs by the Drosophila segmentation gene network. *Proc. Natl. Acad. Sci. U. S. A.* **110**, 6724-6729 (2013).
26. U. Löhr, H.-R. Chung, M. Beller, H. Jäckle, Antagonistic action of Bicoid and the repressor Capicua determines the spatial limits of *Drosophila* head gene expression domains. *Proc. Natl. Acad. Sci. U. S. A.* **106**, 21695-21700 (2009).
27. V. Irish, R. Lehmann, M. Akam, The *Drosophila* posterior-group gene *nanos* functions by repressing *hunchback* activity. *Nature* **338**, 646-648 (1989).
28. M. Hülskamp, C. Schröder, C. Pfeifle, H. Jäckle, D. Tautz, Posterior segmentation of the *Drosophila* embryo in the absence of a maternal posterior organizer gene. *Nature* **338**, 629-632 (1989).
29. M. D. Petkova, G. Tkačik, W. Bialek, E. F. Wieschaus, T. Gregor, Optimal decoding of cellular identities in a genetic network. *Cell* **176**, 844-855.e815 (2019).
30. S. Surkova *et al.*, Quantitative dynamics and increased variability of segmentation gene expression in the *Drosophila Krüppel* and *knirps* mutants. *Dev. Biol.* **376**, 99-112 (2013).
31. E. D. Eldon, V. Pirrotta, Interactions of the *Drosophila* gap gene *giant* with maternal and zygotic pattern-forming genes. *Development* **111**, 367-378 (1991).
32. G. Struhl, P. Johnston, P. A. Lawrence, Control of *Drosophila* body pattern by the *hunchback* morphogen gradient. *Cell* **69**, 237-249 (1992).
33. R. Kraut, M. Levine, Mutually repressive interactions between the gap genes *giant* and *Krüppel* define middle body regions of the *Drosophila* embryo. *Development* **111**, 611-621 (1991).
34. X. Zhu *et al.*, Speed regulation of genetic cascades allows for evolvability in the body plan specification of insects. *Proc. Natl. Acad. Sci. U. S. A.* **114**, E8646-E8655 (2017).
35. K. R. Wotton *et al.*, Quantitative system drift compensates for altered maternal inputs to the gap gene network of the scuttle fly *Megaselia abdita*. *eLife* **4**, e04785 (2015).
36. E. El-Sherif, J. A. Lynch, S. J. Brown, Comparisons of the embryonic development of *Drosophila*, *Nasonia*, and *Tribolium*. *Wiley Interdiscip. Rev. Dev. Biol.* **1**, 16-39 (2012).
37. Manu *et al.*, Canalization of gene expression and domain shifts in the *Drosophila* blastoderm by dynamical attractors. *Plos Comput. Biol.* **5** (2009).
38. M. J. Pankratz, M. Hoch, E. Seifert, H. Jäckle, *Krüppel* requirement for *knirps* enhancement reflects overlapping gap gene activities in the *Drosophila* embryo. *Nature* **341**, 337-340 (1989).
39. M. Mlodzik, G. Gibson, W. J. Gehring, Effects of ectopic expression of *caudal* during *Drosophila* development. *Development* **109**, 271-277 (1990).
40. R. Rivera-Pomar, X. Lu, N. Perrimon, H. Taubert, H. Jäckle, Activation of posterior gap gene experssion in the *Drosophila* embryo. *Nature* **376**, 253-256 (1995).
41. S. Han, J. Pool, J. Tran, W. J. Dally, Learning both weights and connections for efficient neural networks. arXiv:1506.02626 (08 June 2015).

42. M. Raissi, Deep hidden physics models: deep learning of nonlinear partial differential equations. *J. Mach. Learn. Res.* **19**, 1-24 (2018).
43. Z. Long, Y. Lu, X. Ma, B. Dong, PDE-net: learning PDEs from data. arXiv:1710.09668 (26 October 2017).
44. T. Qin, K. Wu, D. Xiu, Data driven governing equations approximation using deep neural networks. *J. Comput. Phys.* **395**, 620-635 (2019).
45. D. Rolnick, A. Veit, S. Belongie, N. Shavit, Deep learning is robust to massive label noise. arXiv:1705.10694 (30 May 2017).
46. R. N. Gutenkunst *et al.*, Universally sloppy parameter sensitivities in systems biology models. *Plos Comput. Biol.* **3**, 1871-1878 (2007).
47. J. F. Apgar, D. K. Witmer, F. M. White, B. Tidor, Sloppy models, parameter uncertainty and the role of experimental design. *Mol. Biosyst.* **6**, 1890-1900 (2010).
48. D. Papatsenko, M. Levine, The *Drosophila* gap gene network is composed of two parallel toggle switches. *PLOS ONE* **6**, e21145. (2011).

**Figure Legends**

**Fig. 1.** Method demonstration with biochemical adaptation of a two-node network. (*A*) The functional dependency of the synthesis terms $f_1$ and $f_2$ for gene 1 ($g_1$) and gene 2 ($g_2$) are evaluated by a DNN (dashed rectangle). In this example, $f_1$ (shaded blue circle) and $f_2$ (shaded green circle) can depend on all the three variables: $g_1$ (solid blue circle), $g_2$ (solid green circle) and the input signal *I* (solid pink circle). Time evolution of the dynamic system corresponds to recurrent iteration of the DNN. The output $\boldsymbol{g}(t)= (g_1(t), g_2(t))$ is compared with the target value to define the Loss function for training. (*B*) We require $g_1$ to be adaptive to the input change (pink line) and set its target value at certain specific time points as indicated by the blue squares. No constraints on $g_2$ are imposed. Blue and green lines are the time evolution of $g_1$ and $g_2$, respectively, after training. (*C*) $f_1$ (blue dotted line) and $f_2$ (green dotted line) as functions of *I*, $g_1$ and $g_2$, respectively, after training. The three panels show their dependence on *I*, $g_1$ or $g_2$ with the other two variables fixed. (*D*) The regulation network drawn from the information of (*C*). For example, the first panel of (*C*) indicates that $f_1$ increases with *I*, implying *I* activates $g_1$.

**Fig. 2.** Manipulating the black box of DNN to gain information about gene regulations. (*A*) Method for blocking the regulation from $g_2$ to $g_1$, i.e. $f_1$ has no direct dependency on $g_2$. (*B*) This technique can be applied after the model is trained to simulate the effect of the mutant in which a specific regulation is deleted. For example, with the trained DNN, blocking the self-regulation of $g_1$ results in an increase in $g_1$ (from lighter to darker blue line), indicating a self-inhibition. Similarly, the effect of blocking the regulation from $g_1$ to $g_2$ indicates a repression from $g_1$ to $g_2$. (*C*) Such blocking can also be applied before training, introducing specific constraints to sparsen the regulation network. With only those regulations in grey unblocked (left panel), the minimal feed-back and incoherent feed forward motifs for adaptation emerge (right panel). (*D*) Sparsening the regulation network without prior knowledge by "mutation and selection". The upper panel plots the sensitivity and adaptation error for the trajectory of networks described in the main text. The lower panel shows two of such networks at the indicated steps. Note that the minimum incoherent feed forward motif appears before the network has too few links to adapt.

**Fig. 3.** Simulating the spatial patterning of the gap genes. (*A*) Left panel: Maternal morphogens form spatial gradients across the embryo, serving as inputs to the system. As an example, the image of the morphogen Bcd tagged with GFP is shown on the top (Reproduced from Ref. (25)). Its quantification is shown in the lower graph (pink curve). The same graph also shows the quantification for the other two morphogens Cad (yellow curve) and Tor (grey curve) (Data from Refs: (26, 37)). Right panel: Gap genes respond to morphogen gradients and regulate each other to form stripe patterns. The top panel shows the image of immunofluorescence staining of the gap protein Kr (Reproduced from Ref. (22)). Its quantification (green) as well as those for the other gap proteins (Hb (blue), Kni (red), Gt (blue)) are shown in the lower panel (Data from Ref. (22); see also *SI Appendix*, S3). Seven snap shots of the time evolution (dynamics) of the gap gene expression is used as training target (Methods). (*B*) Dynamics at different space points are independent. At each spatial point *x*, the model has the same structure as in the previous adaptation example. (*C*) The dynamics of the trained model reproduces the WT data with high accuracy. Shown are the model fittings (solid lines) to the data (dotted lines) for the first and the last snap shots used in training. See *SI Appendix*, S4 for fitting results of the rest snap shots.

**Fig. 4.** Model predictions on gap-gene patterns in different mutants. (*A*) Examples of model prediction on mutant pattern (upper panel, solid line), in comparison with the corresponding experimentally measured profiles (lower panel, dotted line) (from Refs. (29, 30)). (See *SI Appendix*, Fig. S4B for more comparisons). (*B*) Predicted feature positions (along the A-P axis) *versus* feature positions from the real data. There are all together 159 features in the model and 166 features in the real data for 13 different mutants. Among those, 152 features matched between model and data (dark grey points around the diagonal). The positional root-mean-square-error (RMSE) for comparison of these points is 4% embryo length (EL).

Features in WT (which are used in the training) are also shown (yellow squares). There are 14 features from the real data that are not predicted by the model (light grey points on the horizontal axis), and there are 7 features predicted by the model that did not show up in the real data (pink points on the vertical axis). Insert: Statistics of the prediction results. (*C*) Both Kr and Hb are long thought to strongly inhibit *gt*, but in *bcd⁻Kr⁻* mutant where both the Kr domain and the central Hb boundary are lacking, Gt domain does not expand much towards the anterior pole (lower panel, picture reproduced from Ref. (32)). Prediction is consistent with this phenomenon (upper panel). (*D*) The *nos⁻* mutant has a severe phenotype (see Fig. 4A, second column). Deletion of maternal Hb in *nos⁻* mutant rescues abdomen stripes, with anterior pattern unaffected, resulting in a predicted pattern almost identical to WT. This rescue is consistent with experiments (27). (*E-F*) The lower panels show Kr expression in two other multi-mutants (32, 33). (*zhb⁻* means mutating zygotic *hb* only, leaving maternal Hb unaffected.) Upper panels show the model prediction. The expression of Kr is the green line.

**Fig. 5.** Uncover the regulation network via link blocking. (*A*) Blocking the effect of Gt on *kni* results in an expansion of Kni domain towards Gt domain, implying Gt inhibits *kni*. Applying this "knockout" one by one to all links between genes yields an interaction network shown in (*B*). Note that anterior and posterior domains of Hb and Gt are known to obey different regulations. Thus, they are treated separately and labeled with the subscripts a and p, respectively. (*C*) The gap-gene network inferred from a large body of experiments (23).

**Fig. 6.** Removal of redundant regulations. (*A*) A successful solution free of redundant activations can be found by blocking the undesired regulations before training, using the same technique as Fig. 2C. Left panel: the network topology used for training. Gray (directional) links indicate possible regulations (without specifying the sign). Right panel: the resulting network after training. (*B*) At each step, delete the most irrelevant gap gene cross-regulation for pattern formation (leaving the maternal inputs always connected), retrain and iterate. Prediction

accuracy decreases as cross regulations are deleted, using the criteria of Fig. 4B. Solid lines: percentage of data features correctly predicted. Dotted lines: mean square positional error between these predicted and data features. False positive rates are not very informative and not shown. Results of six repeated runs are basically the same. Average networks (using majority rule) along this track are shown on the right, from redundantly connected to over simplified. There exists a point where redundancies are largely removed while core topology is basically kept (when 11 links are removed).

Fig. 1

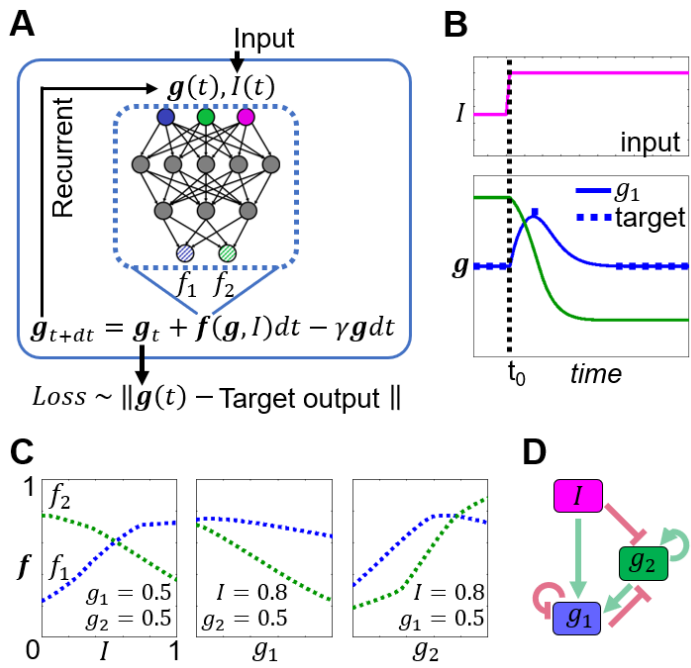

Fig. 2

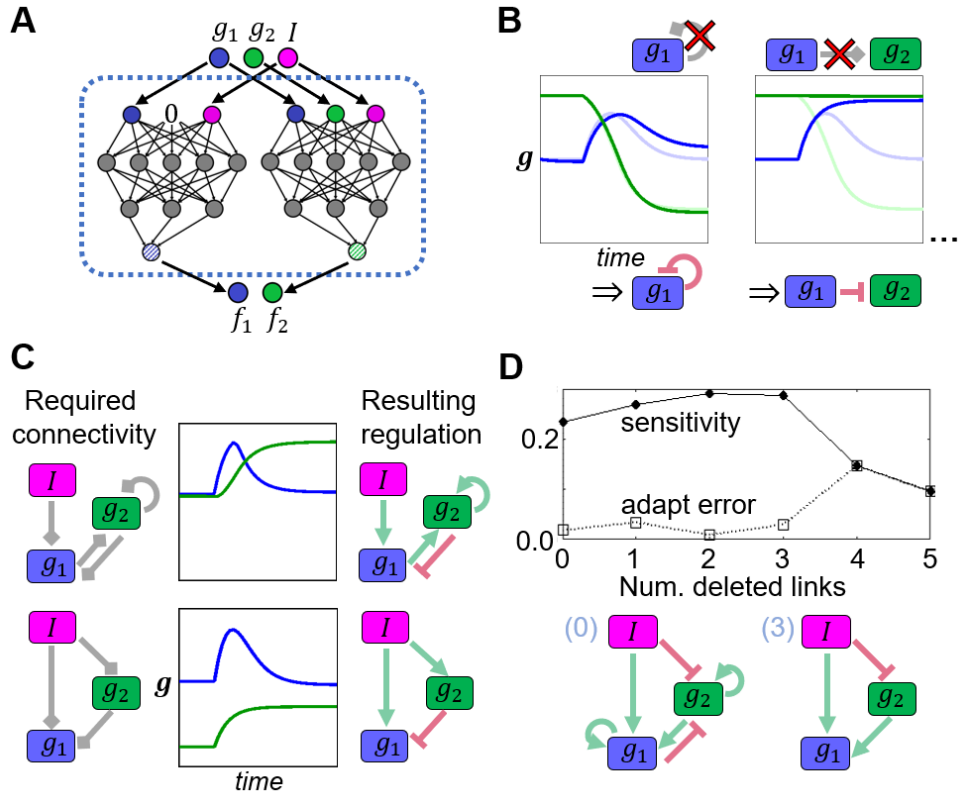

**Fig. 3**

## A

**Maternal morphogens**
e.g. Bicoid (Bcd):

**Gap genes**
e.g. Krüppel (Kr):

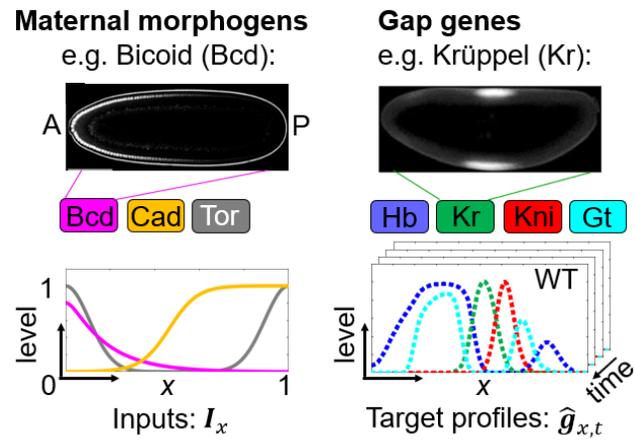

Inputs: $I_x$

Target profiles: $\hat{g}_{x,t}$

## B

state $t$:

$$g_{x,t+dt} = g_{x,t} + f(g_{x,t}, I_x)dt - \gamma g_{x,t}dt$$

state $t+dt$:

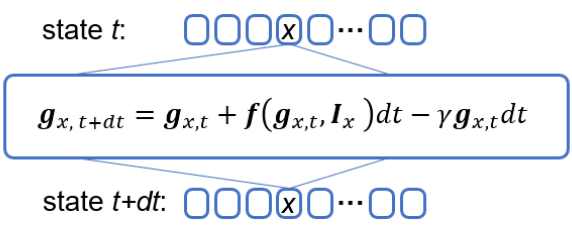

## C Fittings:

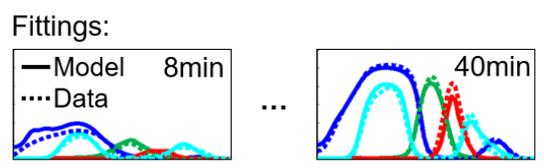

**Fig. 4**

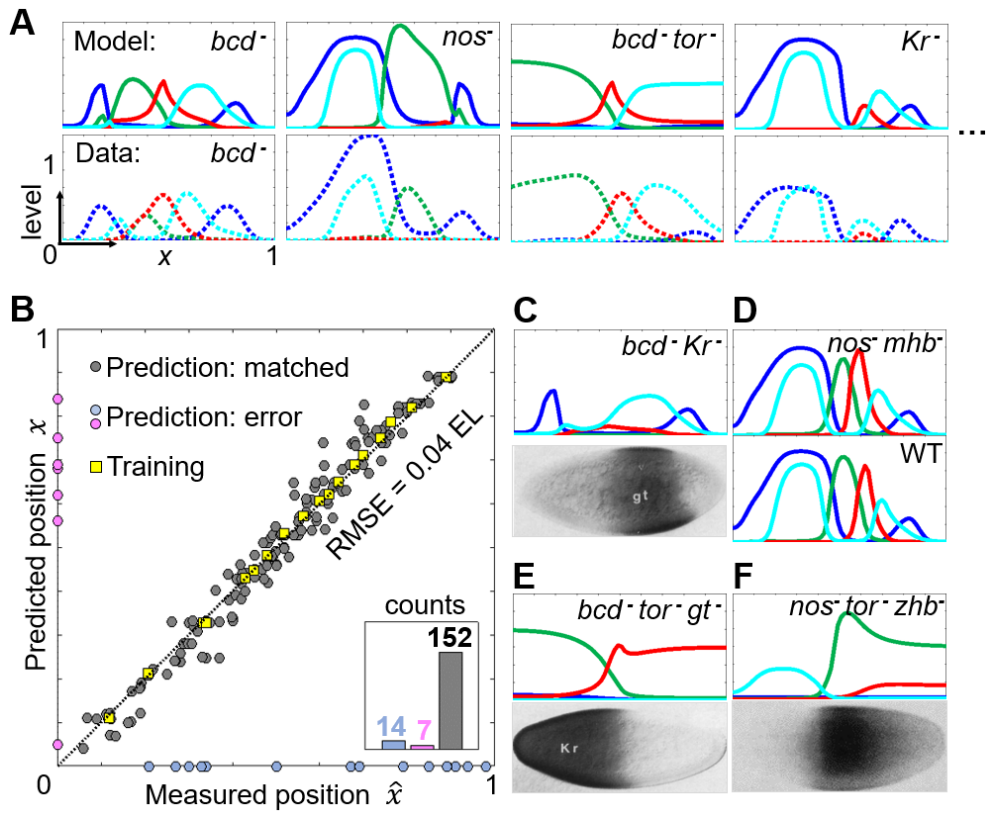

**Fig. 5**

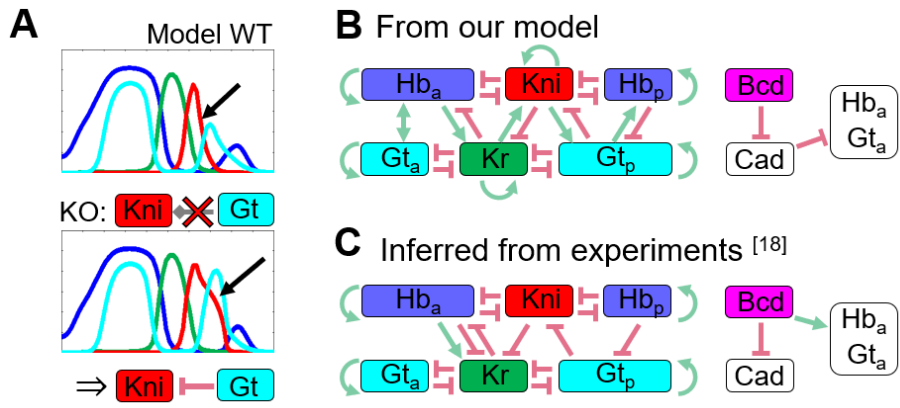
30

**Fig. 6**

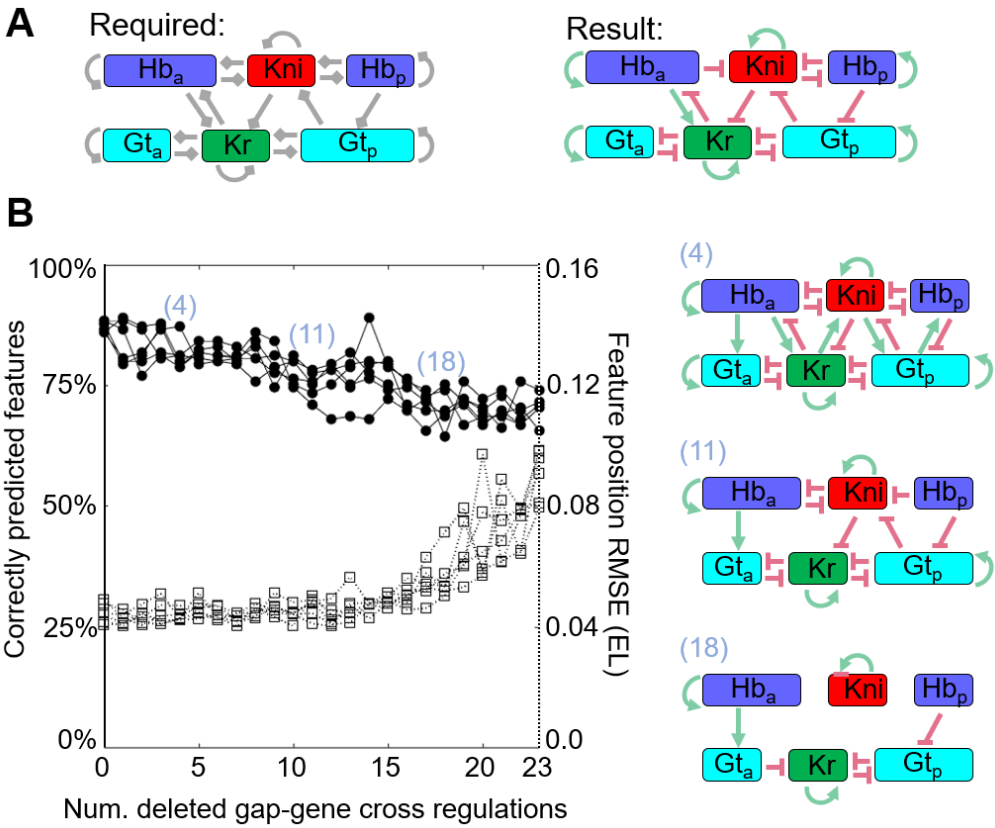

# Supplementary Information for

**Deciphering gene regulation from gene expression dynamics using deep neural network**

Jingxiang Shen, Mariela D. Petkova, Yuhai Tu, Feng Liu and Chao Tang

Correspondence to: tangc@pku.edu.cn

**This PDF file includes:**

Supplementary text S1 to S10

Figures S1 to S10

Tables S1 and S2

SI References

**(S1) The Adaptation task**

**Definition of loss function.**

Dynamics are simulated following Fig. 1B in the main text. Input level is 0.5 before $t_0$ ($t_0$=5 timesteps), and a random number larger than 0.5 (the right half of a normal distribution centered at 0.5, standard deviation=1) after $t_0$ for each training step. Initial condition: $g_1\_initial$=0.5, while $g_2\_initial$ is a tunable variable.

Loss function consists of four parts. $Loss_1$ requires the starting point to be a fixed point:

$$Loss_1 = \sum_{t=1}^{5}(g - g_{(initial)})^2$$

Then input jumps from 0.5 to a higher level.

After 3 timesteps, $Loss_2$ requires node $g_1$ (output node) have a response of 0.8 at $t=8$.

$$Loss_2 = 5 * (g_1 - 0.8)^2$$

Then the dynamics given 10 timesteps to "adapt", before $Loss_3$ and $Loss_4$ are applied: $g_1$ is required to relax back to its initial value (0.5) by $Loss_3$:

$$Loss_3 = \sum_{t=19}^{24}(g_1 - 0.5)^2$$

And the "adapted" state should also be a fixed point:

$$Loss_4 = \sum_{t=19}^{24} (g - g_{(t=18)})^2$$

**Plotting the learnt regulation function.**

Though DNN itself is complex, the 2-dimentional regulation function *f* can be plotted directly. (Fig. S1A)

**Enumerating all possible connectivity topologies.**

With 6 possible edges (from *g₁*, *g₂*, *I* to *f₁*, *f₂*), there are $2^6$ possibility in total. Among them, topologies with no connection (direct or indirect) from Input to *g₁* (output), and topologies with redundant node (some edges links to *g₂*, but *g₂* have no link to output) are not considered. Training repeated four times for each connectivity topology. Following Ref. (1), we use peak response (sensitivity) and adaptation error (precision) to quantify different solutions. A bunch of solutions with different number of regulation edges could be found to achieve adaptation (Fig. S1C). Among them, exist minimal networks of negative feedback loop and incoherent feed forward loop (Fig. 2C main text). Note that there also emerge three topologies with non-monotonic regulations, e.g. *g₂* can activate *g₁* at low level but repress *g₁* at higher level, then node *g₁* performs adaptation if *g₂* increase monotonically by activation from input. (Fig. S1B)

**Perfect adaptation by a feed-back loop.**

If *g₁* retains precisely its initial value ($Loss_3$=0), the system is called to achieve perfect adaptation. It is known (from analyzing linearized equations near the fixed points) that it is impossible for topology like Fig. S1D to achieve perfect adaptation, but for that of Fig. S1E (identical to upper panel in Fig. 2C) it is possible (2). This result can also be explained intuitively: for network of Fig. S1D, *g₂* is only determined by *g₁*, so if *g₁* adapts perfectly, *g₂* goes back to its initial value, leaving the change in input cannot be canceled on *g₁*, thus *g₁* cannot adapt perfectly. This argument is quite general thus applies to our DNN based model as well. Figs. S1D-E also suggest that DNN will achieve perfect adaptation whenever it is possible.

**(S2) Direct regularization does not help sparsening the regulation network**

Training with weight decay makes the DNN sparser. But sparsity in the deep neural network does not mean sparsity in the effective regulation network that it represents, as parameters in DNN do not have explicit correspondence to edges in the regulation network. So, the role of blocking regulations explicitly at the effective regulation network level (Fig. 2A of main text) cannot be replaced by simple normalization techniques. To justify this, we did not block any of the edges, and see what the L2 normalization can achieve. The strength of L2 normalization term (rate of weight decay) is increased from zero to a very large value, so large that the network can no longer be trained (Fig. S2). We found that network #17, where weight decay is extremely strong, has the same topology as network #1. Besides, all of the successful solutions (small training loss, also good in the view of peak response and adaptation error) have all six regulation edges, no matter how strong the normalization is.

**(S3) A brief introduction to gap-gene network**

**Maternal morphogens.**

Positional information along A-P axis is thought to derive solely from gradients of maternal effect morphogens: anterior Bicoid (Bcd), posterior Nanos (Nos), and Torso (Tor) at both termini. The Bcd gradient also shapes maternal component of Caudal (Cad) protein by translational repression of uniformly distributed *cad* mRNA. The repression is fitted explicitly form measured Bcd and Cad profiles (3-5) as a Hill function:

$$\text{Cad} = 0.0008/(0.0008 + \text{Bcd}^{2.5}) \qquad (S.1)$$

Note that Cad also have zygotic component (repressed in the anterior half by maternal Hunchback), but maternal and zygotic Cad are redundant functionally (6), so only maternal component is considered here. As a piece of knowledge, the only effect of Nos is taken to be repressing maternal Hunchback (mHb) in the posterior half (7, 8), thus determining initial condition for the gap-gene dynamics in our model setting. Pattern of Torso is a little bit complicated, as Tor protein actually uniformly distribute throughout the embryo and is only activated by cleaved Trk secreted into the perivitelline space meditated by Torsolike (9). It is difficult to measure directly Tor kinase activity. Measurements in Ref. (10) indicates that the activity of Tor-Response-Element can be detectable within the 23% EL region at both termini. Protein pattern of Capicua (11), which is repressed by Tor, also confirm this result. So, in our model, the pattern of Tor (at both termini) is treated as half-gaussian shape with std=0.15. It naturally drops below 0.1 of its peak value at 0.23 EL away from the pole. Several factors (especially terminal gep genes Tailless (Tll) and Huckbein (Hkb)) acts at down streams of Tor, which are not explicitly considered in our model.

**Remove irrelevant features in gap-gene pattern.**

With the maternal morphogens as inputs, gap-genes (usually referred to the four trunk gap genes), *hunchback* (*hb*), *Krüppel* (*Kr*), *knirps* (*kni*), and *giant* (*gt*), form 10-20 nuclei wide expression domains. It is important to note that certain features are known to be caused by regulations outside the gap-gene network, and are largely decoupled from the rest of the gap-network. For *kni*, its anterior cap domain is activated by Dorsol (together with Bcd), and the narrow anterior stripe domain is activated by Buttonhead (12). These two factors belong to the dorsal-ventral system (instead of the A-P system considered here) and head patterning system. For *gt*, its anterior most small domain and the cleavage of anterior major expression domain (at around 30% from anterior to posterior, due to *empty spiracles* (13)) appear only after late n.c. 14 and sharpen during cellular blastoderm stage (13, 14). This temporal evidence and the strong dependence to head gap-genes suggests that they are decoupled from pattering of the trunk segments. So, these features are removed manually in the patterns on which DNNs were trained (see the "target profile" in Fig.3A of main text, and Fig. S3C). These features are also not considered when comparing measured mutant patterns with model predictions.

**Ignoring diffusion.**

Gap-gene pattern forms during blastoderm stage, where the gap-gene network follows reaction-diffusion dynamics basically, but diffusion is ignored in our model mainly for simplicity. And we have actually tried to train the model with diffusion (with fixed or tunable diffusion constant), and no significant improvement can be observed. Note that Reinitz et.al. have pointed out long ago that though diffusion exists, they are not playing any essential role in the gap-gene regulation mechanism (15).

**Wild-type and mutant expression profiles.**

We chose published data for gap gene profiles: time course data in n.c. 14 of wild-type (WT) from Gregor lab (16), a single snapshot at around 40 minute into n.c. 14 of the maternal factor mutant from Gregor lab (17), certain gap gene mutant ($Kr^-$, $kni^-$, and $Kr^-kni^-$) from Reinitz lab (18, 19), and semi-quantitative data of other gap gene mutants ($hb^-$ & $gt^-$) from papers in the 1990s (13, 20-22). All these are protein data, obtained via immunostaining on fixed embryos.

**Data pre-processing scheme.**

Since the data obtained with different experiment methods, from different labs, or even different batches may bear large systematic deviations from each other. We further processed the collected data to keep them self-consistent. Data pre-processing are organized as 6 steps: (1) we only use mean expression profile of many embryos at the same temporal stage, as subtle subjects as noise attenuation is beyond the scope of this work. (2) Expression level of the four gap genes are normalized according to their spatial maximal in WT pattern at 40 minutes into n.c. 14, just as Ref.(17) (except for $hb^-$ & $gt^-$). (3) A spatial-temporal Gaussian filter is applied to smooth the time course data. Standard deviation of the Gaussian kernel is set to be 2% embryo length spatially and 5 minutes temporally. (4) While most profiles are extracted from the dorsal side of the mid-sagittal plane of unflatten embryos, profiles of $Kr^-$, $kni^-$, and $Kr^-kni^-$ double mutant were obtained from the middle surface of slightly flattened embryos. By comparing WT pattern measured via these two methods, a nonlinear transformation in positional coordinates $x$ should be applied to makes these three profiles have consistent coordinate system with the others. (5) Peak heights measured in these three flattened embryos (*Kr⁻*, *kni⁻*, and *Kr⁻kni⁻*) are also adjusted. As peak heights in WT profile measured with the same method have systematic deviation from the dorsal patterns. (6) For *hb⁻* & *gt⁻*, profiles are extracted from stained embryo images from papers published in 1990s (13, 14, 20, 22), both time point and embryo orientation are not carefully controlled, and no normalized factor is available (simply normalized with their own peak values); so only number of expression peaks and their rough positions make sense.

### (S4) Training and predictions

**The training set and fitting.**

Training data (dashed line in Fig.S4A) consists of a time series of WT gap-gene patterns (first 40 minutes, n.c. 14), snapshot pattern (at ~40 min into n.c. 14) of two mutants lacking all maternal morphogens (*bcd⁻nos⁻tor⁻* and *bcd⁻nos⁻tor⁻mhb⁻*). A typical fitting is presented at the same time (as solid lines). Gap-gene profiles of *bcd⁻nos⁻tor⁻* have been measured quantitatively in Ref.(17), where Kr is uniformly high. This is consistent with the anterior half of *bcd⁻tor⁻* (also Kr uniformly high), as Nos only presents in the posterior half. As for the pattern of mutant *bcd⁻nos⁻tor⁻mhb⁻*, we know from Ref.(23) that Gt is uniformly high. This is also consistent with the posterior half of *bcd⁻tor⁻* as Bcd naturally does not present in the posterior half. Its expression value is *assumed* to be in consistent with this mutant, as no direct quantitative measurement is available.

**Settings of the loss function.**

Loss function for Figs. 2 to 4 in the main text consist of two parts. Euclidian distance between simulated and measured profiles of wild-type time series:

$$Loss_1 = \sqrt{\sum_{t,x}(g_{WT}(x,t) - \hat{g}_{WT}(x,t))}$$

And squared Euclidian distance on the two mutant profiles at the final time step (40 minutes into

n.c. 14)

$$Loss_2 = \sum_x \left(g_{mut}(x, t_{end}) - \hat{g}_{mut}(x, t_{end})\right)$$

The reason for using different norms with the two losses is that as the mutant profiles are not accurate (especially not guaranteed to be consistent with WT), we hope them to help guiding the optimization towards the right direction at the beginning ($Loss \approx 1$), and fade away gradually (when $Loss \ll 1$), leaving WT data to finally refine the solutions. A weak weight decay term, and an explicit Langevin noise term are also added.

**Predictions on mutants with (semi-)quantitative measured profiles.**

The number, position and even the relative intensity of almost all the peaks in the gap gene profiles of a number of mutants are well predicted. Interestingly, some delicate details are captured successfully by prediction: for example. in *Kr⁻* mutant, Kni peak changes position and lies under Gt peak. And when Bcd dosage is halved or doubled (*bcd*1X or *bcd*4X), the predicted posterior boundary of the anterior Hb domain shifts by -8% or +9.3%, which is very close to the experimental measured value of -6.5% or +9.4% (4), rather than ±11.6% as predicted by a simple threshold activation model.

**Quantification of the predictions.**

To compare the predicted mutant patterns against the experiment measured ones, features (peaks and rising/falling boundaries) in these patterns are extracted and matched between the predicted and data profiles as Fig. 4B in the main text. The comparisons are made at around 40 minutes n.c. 14. (Step 1) *Extracting features*: Only peaks higher than 0.17 (normalized expression level) are considered. For a peak *i* (position denoted as $p_i$), its nearest rising and falling boundaries (defined by half-height of peak *i*) are found (positions: $b^{(rising)}_i$ and $b^{(falling)}_i$). After locating all peaks of a certain gene, certain boundaries found in this way would appear to be redundant. For example, if peak *i* is to the left of peak *j* ($p_i < p_j$) but the rising boundary of peak *j* is on the left side of peak $i$, ( $b^{(rising)}_j < p_i < p_j$) then $b^{(rising)}_j$ is no longer considered as a "feature". (Step 2) *Matching features*: Two features of the same gene, one from predicted pattern and the other from data, are matched if both of them are the nearest to the other compared with all other features of the same type (peak/rising/falling). Gap-gene patterns of 13 mutants (those shown in Fig. 4B, plus *Kr⁻kni⁻* (19)) as well as WT are used for this comparison.

**(S5) Interpreting the regulation logic**

**The link blocking technique.**

In main text Fig. 2A we have presented the method of blocking a single regulation edge from one node to another, in order to obtain the gene regulation network learnt by the DNN. The procedure of doing this is explained in more details here, with the case of Fig.5 in the main text:

$$(f_1, f_2, f_3, \cdots, f_n) = F(x_1, x_2, x_3, \cdots, x_m)$$

$$(\tilde{f}_1, \tilde{f}_2, \tilde{f}_3, \cdots, \tilde{f}_n) = F(0, x_2, 0, \cdots, x_m)$$

The desired output is thus $(f_1, \tilde{f}_2, f_3, \cdots, f_n)$. We can thus integrate the pattering process with this modified "regulation function", yielding Fig.5A, from which one can see the expansion of contraction of certain expression domains. A quantitative measure is defined to extract corresponding regulation logic from such expansion of contraction. A "WT" pattern (profiles of each gap-genes along the A-P axis) is calculated first for reference:

$$\{u_n(x)\} \quad n = 1,2,3,4$$

A "mutant" pattern with interaction from gene $i$ to $j$ being blocked forms a different pattern:

$$\{v_n^{(i,j)}(x)\} \quad n = 1,2,3,4$$

Effect from $i$ to $j$ is defined as "the difference of $j$ due to $i$":

$$W_{i \, to \, j} \equiv -\int_{x=0}^{1} \left(v_j^{(i,j)} - u_j\right) u_i \, dx$$

When $j$ domain expands toward the $i$ domain when not "seeing" $i$, $\left(v_j^{(i,j)} - u_j\right)$ is positive where $i$ domain sits, the inner product is positive, thus $W_{i \, to \, j}$ is negative. Repeating this procedure for all pairs of *ij* result in Fig.5-B.

**Strong self-activation is an artifact.**

The only commonly accepted self-activation among the gap-genes are that of *hb* (whose strength is still not quite clear). But in our DNN model, all four gap-genes rely heavily on self-activations. We think one important reason for this artifact is that we have ignored the distinction between mRNA and protein. Here expression data are all protein profiles during n.c. 14, but in reality, gap-gene mRNAs have already accumulated during n.c. 13 (under regulations of the maternal morphogens), thus the rapid accumulation of gap-gene proteins early n.c. 14 is mainly due to purely translation. Without including such effect, another source of activation is needed by the model. Besides this, there may exists some global activating factors (such as Zelda) not taken into the model.

Note that some extra activation is definitely needed for fitting the data, though it may not necessarily be self-activation, as DNN can still find successful solutions if all self-activations apart from Hb are blocked before training. Prediction accuracy drops under these constrains: fraction of correct predicted features drops to 87.4%, and mean square error increases to 5.2% EL (for the intact model of Fig.4 in main text, these two metrics are 91.6% and 4% EL). This result holds if self-activation of Hb is blocked too (training not affected, prediction metrics 84.9% and 5.4% EL).

**The full gap-gene regulation network.**

For the learnt gap-gene network, only the gap-gene cross regulations are presented in Fig. 5B in the main text. We present the whole regulation network, both gap-gene cross regulation and effects of maternal morphogens on gap-genes in Fig. S5A. These are results of 64 repeated trainings.

**Comparison with other methods of interpretation.**

We have tried to visualize the regulation mechanism learnt by DNN with other more conventional methods besides the "link blocking" technique introduced in the main text.

Various methods have been established to visualize convolutional networks. Like the idea of *Feature Visualization* (24), that regulations on a certain output (or hidden) node could be represented by the input vector that maximizes this output (or hidden) node. Or one can construct "Class Activation Map" to see which input features are most relied on when DNN is doing classifications (25). These ideas can be easily transferred to our small fully-connected DNN (7-input, 4-output): for output node $j$, enumerate (with a given step size) all possible inputs combinations (7-d vector, value from 0 to 1), find the 1% (denote by $\text{argmax}'$) that gives the largest value on node $j$, average them, yielding a "mean activation vector" $c_1$; and in the same way find the "mean inhibition vector" $c_2$. Regulations on node j can thus be defined as $c_1-c_2$. Clearly, such "average" operation among the 1% would automatically push certain irrelevant dimensions towards 0.5, which cancels out when $c_1-c_2$ (Fig.S5B, "Activation map").

$$w_j = \underset{g,I}{\text{argmax}'} f_j - \underset{g,I}{\text{argmin}'} f_j \qquad j = 1,2,3,4$$

This procedure works well generally, but as pointed out in the main text, anterior and posterior expression domains of Hb and Gt should better be treated separately. If ignoring this aspect, the regulation network is quite consistent with results in Fig.S5A, from which Fig. 5B in the main text is derived.

Another way of visualizing the regulations can be carried out as follows. 1) Record all inputs and outputs of the DNN, it is of course some nonlinear function. 2) Try to fit it by a linear function, i.e. do linear regression between (*g*,*I*) and *f*. This yields Fig. S5C. Or, 2') one can calculate correlation function between (*g*,*I*) and *f* directly, yielding Fig. S5D. These results bear some differences with Fig. S5 in some details.

**(S6) Important and redundant edges in the learnt gap-gene network**

**No single edge is irreplaceable.**

To search for an important regulation edge in the gap gene network, one can try building up models without it and see if the model breaks down. Network connectivity can be constrained following Fig.2C&D, especially blocking a single edge. So, we can try to find some vital edges, without which the model fails. But results are negative: the model can always fit successfully with any single regulation edge blocked. Moreover, predictions on mutant patterns are also not affected (Fig.S6 first row). Same negative results turned up for any combination of two edges (Fig.S6 second row). It is hard to conclude whether this property originates from the regulation network or deep neural network, or both.

**Removing redundant regulations greedily.**

We have demonstrated in main text (Fig. 6B) that the regulation network can be sparsened if the "least important" edge is removed each time after training, and then the model is retrained with the updated connectivity map. Only gap-gene cross regulations are considered in Fig. 6B, while connectivity from the maternal inputs are kept unchanged. Here we explore the alternative setting, that all edges, including those from maternal inputs, are dropped one by one following this procedure. There are 42 edges in total. Again, we see error increases with more regulations deleted, and prediction accuracy drops all the way towards zero. Some typical sparsened network topologies along this track is also presented below.

**(S7) DNN size and architecture matters little**

To demonstrate that our approach is not sensitive to specific settings of the DNN, we train the model with DNNs of different depth, width, and architectures. All these different network structures have similar performances (Table S1). Among them, the dense-net architecture seems to be the best: fast convergence, good generalization, and insensitive to hyper parameters. But the advantages are not that remarkable. So, in the main text, we use the fully connected multilayer perceptron basically for technical simplicity.

**(S8) Too much or not enough training data**

In the main text we have claimed that including more mutant patterns for training usually brings more noise than informative constrains. Fig.S8A serves as a demonstration, where patterns of three maternal factor single mutants *bcd-*, *nos-*, and *tor-* are also used for training besides those "standard" training set listed in Fig.S4A. We call these mutant patterns *redundant* for training because they can be naturally predicted as in Fig.4 of main text. Without much new information, these data are actually even misleading, leading to even worse predictions on double mutants like *bcd-;tor-* and *nos-;tor-* compared with Fig.S4B.

On the other hand, non-redundant training data helps excluding certain solutions that would otherwise be indistinguishable if the model is constrained with WT data only. In Fig. S8B, the model is trained with WT (Loss$_1$ in section S4) only. It can simulate WT gap-gene dynamics successfully with somewhat "equivalent" but different underlying regulations: Bcd, Nos, and Tor in the model no longer correspond to Bcd, Nos and Tor in reality, as patterns are different when these factors present alone (mutant *bcd-;tor-*, *nos-;tor-*, and *bcd-;nos-*). But DNN stays robust with such functional substitution: it does not affect predicting gap-gene mutants where all maternal morphogens functions as a whole. This is quite a general aspect for black-boxes.

**(S9) Model lacking *kni***

The DNN based model is insensitive to missing important nodes, whose roles can be effectively absorbed into the rest of regulations. A 3-gap-gene model, lacking *kni* completely, can also fit and predict patterns of the three remaining genes. Regulation network is also very similar to the "real one" (Fig. S9).

**(S10) A fragile mechanism is hard to lean**

We have claimed in the main text that DNN can only learn "robust" mechanisms easily. As a negative example, we took an obviously oversimplified gap-gene network topology (Fig.S10A, similar to that proposed in Ref.(26)) as "ground truth", and write down differential equations with Hill-function. For each gap-gene, activation terms are assumed additive, and repressions are multiplicative.

$$\frac{dg_i}{dt} = \left( \sum_j \frac{1}{1 + \left(\frac{K_{ji}^{(+)}}{g_j}\right)^{n_{ji}^{(+)}}} \right) \left( \prod_j \frac{1}{1 + \left(\frac{g_j}{K_{ji}^{(-)}}\right)^{n_{ji}^{(-)}}} \right) - \gamma g_i$$

Parameters are tuned manually (as Table S2).

As a model to generate stripe patterns, the model is, actually, overfitting to the gap gene data. Due to the oversimplified topology, Hill coefficients and interaction strength need to be fine-tuned. Serving as "overfitting ground truth", this model generates a set of synthesized wild-type and mutant patterns (Fig.S10A). Then we trained DNN with exactly the same setting as Fig.3&4 in main text for 8 repeats. Fitting is perfect (Fig.S10B, WT), as well as most of the mutant patterns (e.g. *nos-;tor-* and *bcd-;nos-*). However, it fails to predict mutant patterns like *hb-* and *Kr-* (Fig.S10B). Reasons are straight forward: though able to generate a WT pattern with 5 stripes like real fruit fly, the artificial ground truth network collapses if *hb* or *Kr* were removed (Fig.S10A). But DNN predictions are somewhat more robust against such deletion: other expression domains still exist though expanded and shifted slightly, which is reminiscent of what happens for *real* fruit fly. The underlying regulation network can be mapped out following method of Fig.5A&B, and turned out to be very different compared with the "ground truth", though fitting and most predictions are almost exact. It is now clear, with this negative example, that results of Fig.4 and especially Fig.5 in main text is non-trivial where both predictions and regulation network are correct.

Making use of this oversimplified white-box model (and independent of DNN), we can also test our "knockout" method; basically, confirm that equations in S5 can really capture the sign of regulation. Such interpretation yields Fig. S10E. Comparing with ground truth parameters listed in Table S2, one can easily recognize that every edge has correct sign (activation / repression / none).

# SI References

1. W. Ma, A. Trusina, H. El-Samad, W. A. Lim, C. Tang, Defining network topologies that can achieve biochemical adaptation. *Cell* **138**, 760-773 (2009).
2. M. Y. Zhang, C. Tang, Bi-functional biochemical networks. *Phys. Biol.* **16** (2019).
3. Manu *et al.*, Canalization of gene expression and domain shifts in the *Drosophila* blastoderm by dynamical attractors. *Plos Comput. Biol.* **5** (2009).
4. F. Liu, A. H. Morrison, T. Gregor, Dynamic interpretation of maternal inputs by the *Drosophila* segmentation gene network. *Proc. Natl. Acad. Sci. U. S. A.* **110**, 6724-6729 (2013).
5. T. Gregor, E. F. Wieschaus, A. P. McGregor, W. Bialek, D. W. Tank, Stability and nuclear dynamics of the *bicoid* morphogen gradient. *Cell* **130**, 141-152 (2007).
6. C. Schulz, D. Tautz, Zygotic *caudal* regulation by *hunchback* and its role in abdominal segment formation of the *Drosophila* embryo. *Development* **121**, 1023-1028 (1995).
7. V. Irish, R. Lehmann, M. Akam, The *Drosophila* posterior-group gene *nanos* functions by repressing *hunchback* activity. *Nature* **338**, 646-648 (1989).
8. M. Hülskamp, C. Schröder, C. Pfeifle, H. Jäckle, D. Tautz, Posterior segmentation of the *Drosophila* embryo in the absence of a maternal posterior organizer gene. *Nature* **338**, 629-632 (1989).
9. T. K. Johnson *et al.*, Torso-like mediates extracellular accumulation of Furin-cleaved Trunk to pattern the *Drosophila* embryo termini. *Nat Commun* **6**, 8759 (2015).
10. U. Löhr, H.-R. Chung, M. Beller, H. Jäckle, Antagonistic action of Bicoid and the repressor Capicua determines the spatial limits of *Drosophila* head gene expression domains. *Proc. Natl. Acad. Sci. U. S. A.* **106**, 21695-21700 (2009).
11. O. Grimm *et al.*, Torso RTK controls Capicua degradation by changing its subcellular localization. *Development* **139**, 3962-3968 (2012).
12. M. Rothe, E. A. Wimmer, M. J. Pankratz, M. Gonzalez-Gaitan, H. Jackle, Identical transacting factor requirement for *knirps* and *knirps-related* gene expression in the anterior but not in the posterior region of the *Drosophila* embryo. *Mechanisms of Developmen* **46**, 169-181 (1994).
13. E. D. Eldon, V. Pirrotta, Interactions of the *Drosophila* gap gene *giant* with maternal and zygotic pattern-forming genes. *Development* **111**, 367-378 (1991).
14. R. Kraut, M. Levine, Spatial regulation of the gap gene *giant* during *Drosophila* development. *Development* **111**, 601-609 (1991).
15. J. Jaeger *et al.*, Dynamic control of positional information in the early *Drosophila* embryo. *Nature* **430**, 368-371 (2004).
16. J. O. Dubuis, R. Samanta, T. Gregor, Accurate measurements of dynamics and reproducibility in small genetic networks. *Mol Syst Biol* **9** (2013).
17. M. D. Petkova, G. Tkačik, W. Bialek, E. F. Wieschaus, T. Gregor, Optimal decoding of cellular identities in a genetic network. *Cell* **176**, 844-855.e815 (2019).
18. A. Pisarev, E. Poustelnikova, M. Samsonova, J. Reinitz, FlyEx, the quantitative atlas on segmentation gene expression at cellular resolution. *Nucleic Acids Res.* **37**, D560-D566 (2009).
19. S. Surkova *et al.*, Quantitative dynamics and increased variability of segmentation gene expression in

the *Drosophila Krüppel* and *knirps* mutants. *Dev. Biol.* **376**, 99-112 (2013).

20. M. Hulskamp, C. Pfeifle, D. Tautz, A morphogenetic gradient of Hunchback protein organizes the expression of the gap genes *Krüppel* and *knirps* in the early *Drosophila* embryo. *Nature* **346**, 577-580 (1990).

21. R. Kraut, M. Levine, Mutually repressive interactions between the gap genes *giant* and *Krüppel* define middle body regions of the *Drosophila* embryo. *Development* **111**, 611-621 (1991).

22. X. Wu, R. Vakani, S. Small, Two distinct mechanisms for different positioning of gene expression borders involving the *Drosophila* gap protein Giant. *Development* **125**, 3765-3774 (1998).

23. G. Struhl, P. Johnston, P. A. Lawrence, Control of *Drosophila* body pattern by the *hunchback* morphogen gradient. *Cell* **69**, 237-249 (1992).

24. D. Erhan, Y. Bengio, A. Courville, P. Vincent (2009) Visualizing higher-layer features of a deep network. (Technical Report: University of Montreal).

25. R. R. Selvaraju *et al.*, Grad-CAM: visual explanations from deep networks via gradient-based localization. arXiv:1610.02391 (07 October 2016).

26. D. Papatsenko, M. Levine, The *Drosophila* gap gene network is composed of two parallel toggle switches. *PLOS ONE* **6**, e21145. (2011).

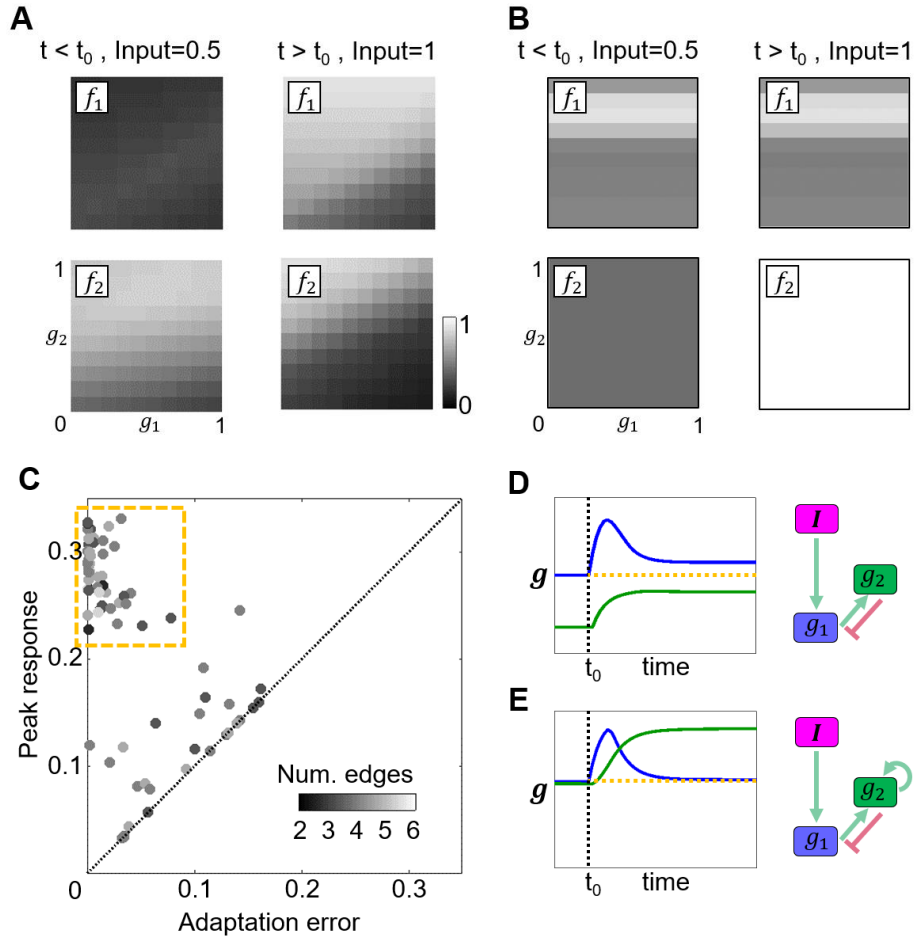

**Fig. S1.** The *adaptation* task. (A) Trained adaptation dynamics presented in Fig. 1B of the main text. Plots of the learnt regulation function, whose intersections are shown as Fig.1C. (B) A 2-edge solution can achieve adaptation relying on non-monotonic interaction. (C) Finding network topologies achieving adaptation. (D&E) Criterion for two-node perfect adaptation (analytical result) (2) also applies to our DNN based models.

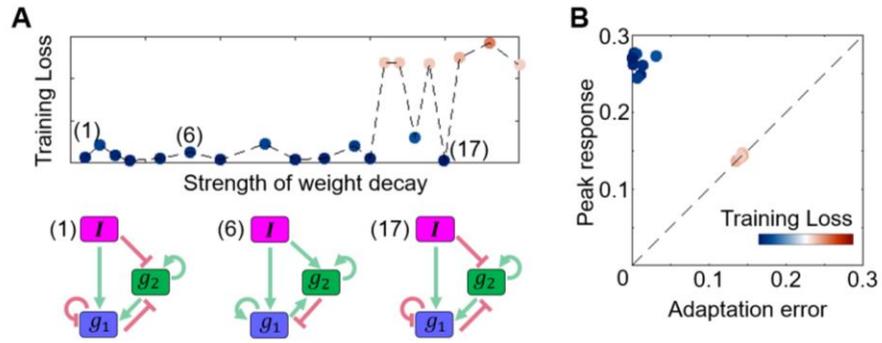

**Fig. S2.** L2 regularization does not help sparsening the effective regulation network. (A) Sparsity is not affected by increasing L2 regularization strength. (B) Small training loss indeed indicates good adaptation.

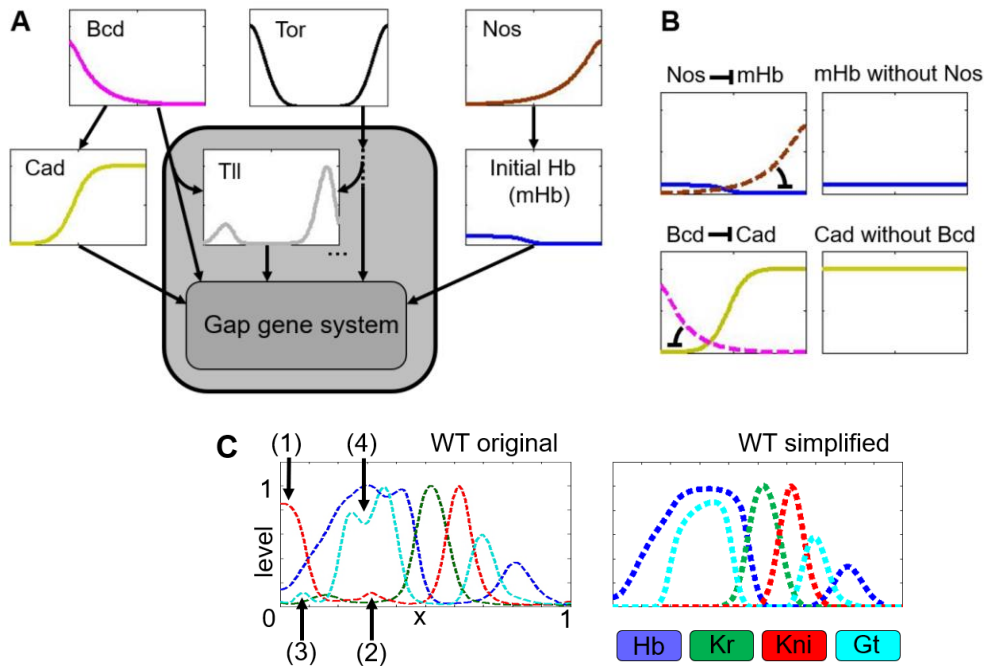

**Fig. S3.** (A) Primary positional information is thought to be carried by three factors: Bcd, Nos and Tor. Other downstream morphogens also play roles in guiding gap gene expression but are themselves regulated by the above three factors, e.g., Cad is repressed by Bcd, mHb is repressed by Nos, Tll is regulated by both Tsl and Bcd. These "downstream" morphogens could in principle be absorbed into the DNN approximately. (B) In *nos*⁻ mutant, profile of mHb is assumed to be a flat line in the whole embryo with the estimated maximum expression level of the mHb in WT. The profile of Cad in *bcd*⁻ mutant is assumed to be a flat line in the whole embryo, with the estimated maximum expression level of the Cad in WT. (C) Remove irrelevant features in gap-gene pattern: anterior cap (1) and anterior stripe (2) domain of Kni, anterior most Gt stripe1 (3), and cleavage in Gt stripe23 (4). Small fluctuations are also removed as backgrounds.

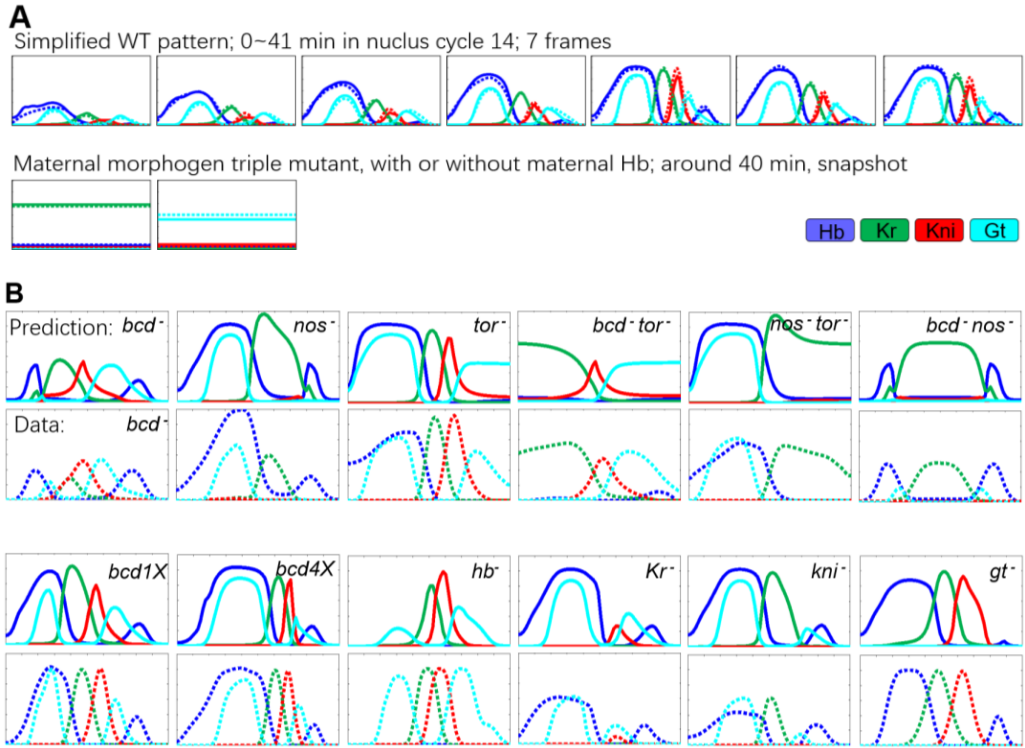

**Fig. S4.** Fittings and predictions. (A) Fitting to training set: training data (dashed lines) consists of 7 frames of WT dynamics, and snapshots of two mutants (*bcd⁻nos⁻tor⁻* and *bcd⁻nos⁻tor⁻mhb⁻*). A typical fitting is presented at the same time as solid lines. (B) Predictions on mutants with quantitative (or semi-quantitative) measured profiles. The predicted expression profiles (solid lines) of Hb, Kr, Kni and Gt along the A-P axis are consistent with the experimentally measured data (dished lines) (4, 13, 14, 17, 19, 20, 22). For *hb⁻* and *gt⁻* data, only the numbers and rough positions of peaks are for comparison due to the semi-quantitative nature of the data.

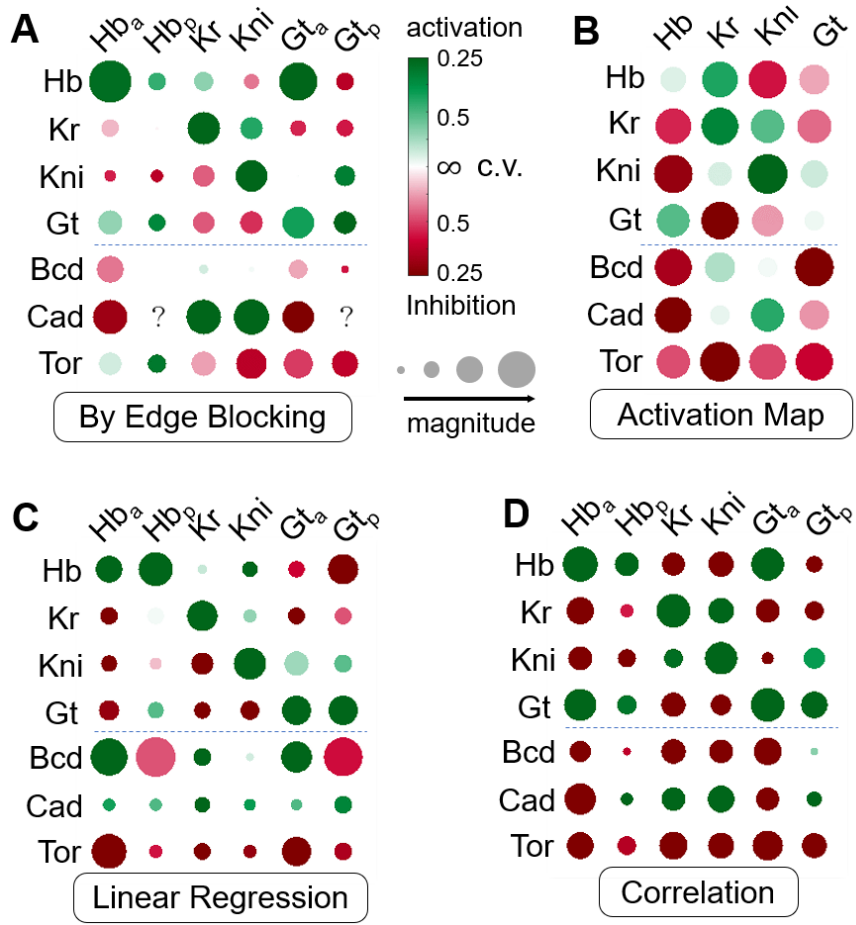

**Fig. S5.** (A) Statistics for the full learnt regulation network of 64 repeated trainings. A sub-network (of the gap gene cross regulations) is shown in Fig. 5B of the main text. Each disc represents the regulation from one gene to another, green for activation, red for inhibition, disc size for regulation strength. Coefficient of variation (c.v.) for each edge is calculated from 64 replica trainings; darker color means smaller c.v. Dashed line separates gap-gene cross regulations and maternal inputs. Question mark means not clear, because anterior domains expand so much making posterior domain indiscernible. (B-D) Regulation network based on activation map, Linear regression, and correlation function. Color bar is the same as A.

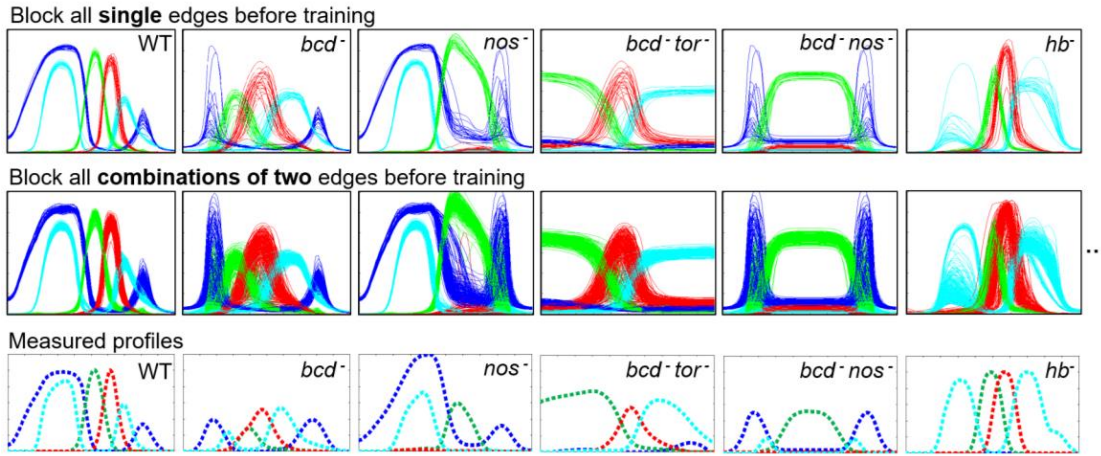

**Fig. S6.** Every regulation edge is not irreplaceable. Fitting (WT) and predictions (others) seemed not to be affected by losing any single (first row) or any combinations of two (second row) regulation edges.

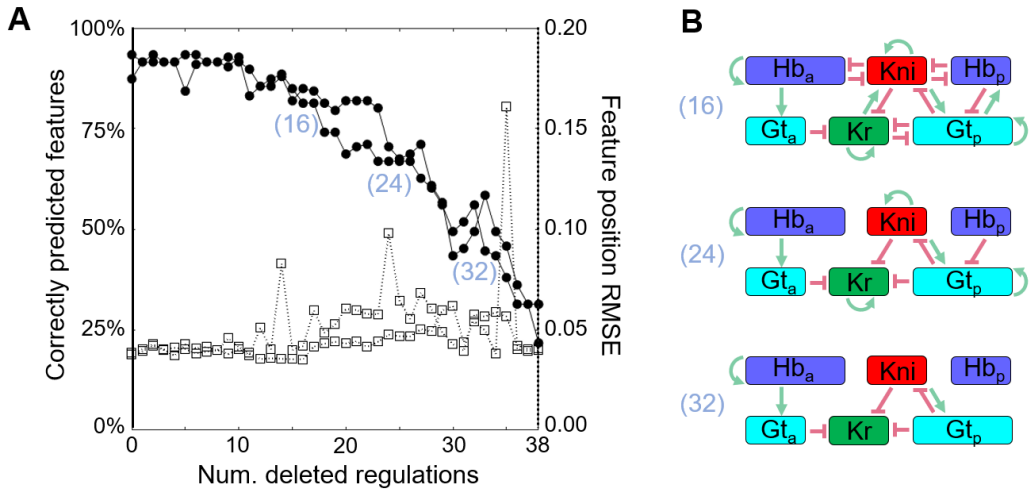

**Fig. S7.** Removing all regulation edges sequentially. The regulation network can be sparsened following the method of Fig.6B. Here all edges, including those from the maternal inputs, are considered.

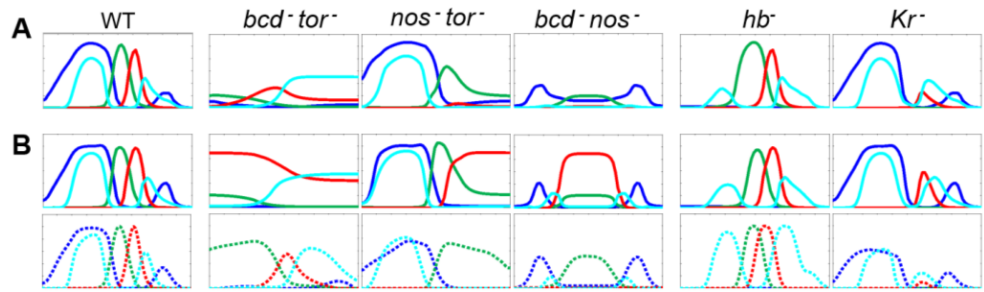

**Fig. S8.** Too much/not enough training data led to over/under-constraining. (A) including another 3 mutant patterns (single mutant *bcd-*, *nos-*, &*tor-*) in training seems to bring more noise than information, as it disturbs predictions on maternal input double mutants. (B) Under-constraining: model trained with WT dynamics only, functions of maternal inputs fused with each other, leading to bad predictions on mutants where these maternal factors function alone. Measured "real" mutant profiles from data are plotted as dashed lines for comparison.

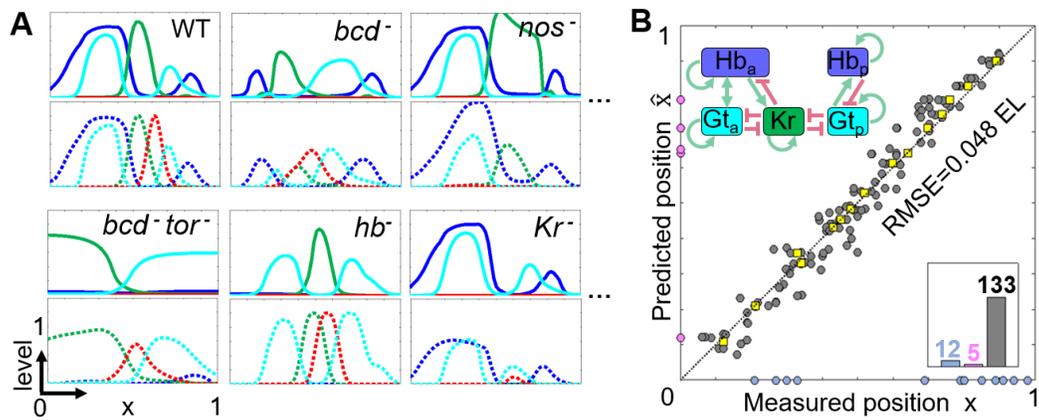

**Fig. S9.** Model still works even without *kni*. (A-B) Pretending *kni* (red) being invisible, a 3-gap-gene model can still fit and predict patterns of the three remaining genes. Regulation network is also very similar to the "real one" (B- upper insert).

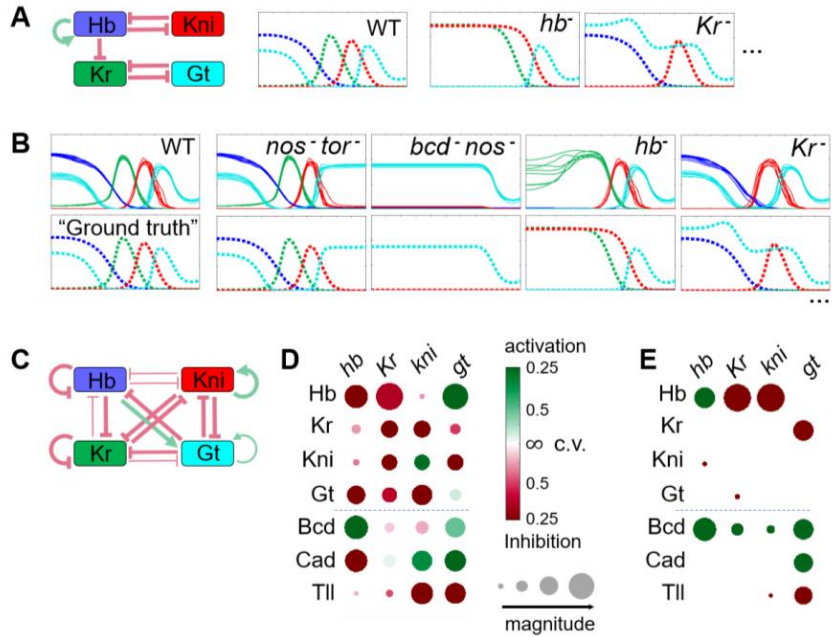

**Fig. S10.** DNN cannot learn an "overfitting ground truth". (A) An oversimplified topology is taken as ground truth and put into differential equations with Hill function. The model generates a set of synthesized data. (B) The DNN model is trained to simulate this data, everything except data is kept unchanged. Fitting is perfect, as well as predictions on some of the mutants, but it failed to predict mutant pattern whenever *hb* or *Kr* is missing. Training is repeated for 8 times. (C&D) The underlying regulation mechanism is very different from the ground truth. (E) Use the white-box model to verify our knockout technique, See below for detail.

**Table S1.** Comparison of different DNN structures with the gap-gene task.

| # | Arch. | Depth | Num. P | Train. Err. | Correct.% | False Positive | Pos. RMSE |
|---|---|---|---|---|---|---|---|
| **FC3** | FC | 3 | 1444 | 0.0054 | 91.6% | 4.4% | 0.040 |
| **FC3_6g** | FC | 3 | 1510 | 0.0045 | 91.6% | 3.8% | 0.039 |
| **FC2** | FC | 2 | 388 | 0.0047 | 93.4% | 4.9% | 0.045 |
| **FC6** | FC | 6 | 4612 | 0.0035 | 90.4% | 8.5% | 0.055 |
| **FC3_w8** | FC | 3 | 172 | 0.0059 | 91.6% | 1.3% | 0.042 |
| **FC3_w128** | FC | 3 | 18052 | 0.0038 | 92.2% | 2.6% | 0.042 |
| **Den3_w16** | Dense | 3 | 1376 | 0.0034 | 90.4% | 2.6% | 0.039 |
| **Den3_w8** | Dense | 3 | 512 | 0.0051 | 92.8% | 5.5% | 0.076 |

**FC**: fully-connected multi-layer perceptron;

**Dense**: densely-connected multi-layer perceptron, where every layer is fully-connected to all pervious layers.

**Depth**: the number of hidden layers+1.

**Num. P**: is the total number of network parameters.

**Train. Err.**: training error on wild-type dynamics (mean Euclidian distance on normalized expression profiles).

**Correct. %**: fraction of measured features that are also predicted.

**False Positive %**: fraction of predicted features that are not present in measured profile.

**Pos. MSE**: positional mean square error of the correct predicted features.

**Table S2.** Parameters for the oversimplified ground truth model for S10. There are 13 regulations in total, (+) means activation and (–) means repression. Each regulation edge has a "strength" 1/K, and a Hill coefficient n.

| Sign | From (j) | To (i) | $1/K_{ji}$ | $n^{(sign)}$ |
|---|---|---|---|---|
| + | Hb | *hb* | 1.5 | 2 |
| - | Kni | *hb* | 1 | 2 |
| + | Bcd | *hb* | 5 | 2 |
| - | Hb | *Kr* | 3 | 4 |
| - | Gt | *Kr* | 1 | 2 |
| + | Bcd | *Kr* | 20 | 3 |
| - | Hb | *kni* | 40 | 2 |
| + | Bcd | *kni* | 80 | 2 |
| - | Tll | *kni* | 5 | 1 |
| - | Kr | *gt* | 40 | 2 |
| + | Bcd | *gt* | 5 | 4 |
| + | Cad | *gt* | 2 | 2 |
| - | Tll | *gt* | 5 | 1 |